\documentclass[twocolumn,english,superscriptaddress,nature]{revtex4-1}

\usepackage[colorlinks,linkcolor=blue,urlcolor=blue,citecolor=blue]{hyperref}
\usepackage{mathtools}
\usepackage{amsmath}
\usepackage{graphicx}
\usepackage{ifpdf}
\usepackage{color}
\usepackage{slashed}
\usepackage{bm}
\usepackage{braket}

\setcitestyle{super}

\begin{document}

\title{Emergent criticality in fully frustrated quantum magnets}

\author{Yuchen Fan}
\affiliation{Beijing National Laboratory for Condensed Matter Physics 
and Institute of Physics, Chinese Academy of Sciences, Beijing, 100190, 
China}

\author{Ning Xi}
\affiliation{Department of Physics and Beijing Key Laboratory of 
Opto-electronic Functional Materials and Micro-nano Devices, Renmin 
University of China, Beijing 100872, China}

\author{Changle Liu}
\affiliation{School of Engineering, Dali University, Dali, Yunnan 671003, 
China}
\affiliation{Shenzhen Institute for Quantum Science and Technology and 
Department of Physics, Southern University of Science and Technology, 
Shenzhen 518055, China}

\author{Bruce Normand}
\affiliation{Laboratory for Theoretical and Computational Physics, 
Paul Scherrer Institute, CH-5232 Villigen-PSI, Switzerland}
\affiliation{Institute of Physics, Ecole Polytechnique F\'ed\'erale 
de Lausanne (EPFL), CH-1015 Lausanne, Switzerland}

\author{Rong Yu}
\affiliation{Department of Physics and Beijing Key Laboratory of 
Opto-electronic Functional Materials and Micro-nano Devices, Renmin 
University of China, Beijing 100872, China}
\affiliation{Key Laboratory of Quantum State Construction and Manipulation 
(Ministry of Education), Renmin University of China, Beijing 100872, China}

\maketitle

{\bf {
Phase transitions in condensed matter are often linked to exotic 
emergent properties. We study the fully frustrated bilayer Heisenberg 
antiferromagnet to demonstrate that an applied magnetic field creates a 
novel emergent criticality. The quantum phase diagram contains four states, 
the DS (singlets on every interlayer dimer bond), DTAF (all triplets with 
antiferromagnetic order), TC (a singlet-triplet checkerboard) and FM 
(saturated ferromagnet). The thermal phase diagram is dominated by a wall 
of discontinuities extending from the zero-field DTAF-DS transition to a 
quantum critical endpoint where the field drives the DTAF and TC into the 
FM. This first-order wall is terminated at finite temperatures by a line 
of critical points, where the Berezinskii-Kosterlitz-Thouless (BKT) 
transition of the DTAF and the thermal Ising transition of the TC also 
terminate. We demonstrate by quantum Monte Carlo simulations that the BKT 
transition does not change the Ising nature of the DTAF-DS critical line. 
By contrast, the combination of symmetries merging on the multicritical 
DTAF-TC line leads to a 4-state Potts universality not contained in the 
microscopic Hamiltonian, which we associate with the Ashkin-Teller model. 
Our results represent a systematic step in understanding emergent phenomena 
in quantum magnetic materials including the ``Shastry-Sutherland compound'' 
SrCu$_2$(BO$_3$)$_2$.}} 


Classical and quantum field theories, formulated to capture the low-energy, 
long-wavelength behaviour of a system, have a foundational role in theoretical 
physics. At a continuous classical or quantum phase transition (QPT), the 
characteristic energy scale vanishes and the correlation length diverges, 
ensuring a profound connection between field theories and the statistical 
mechanics of critical phenomena.\cite{Zinn-Justin2002} In both situations, 
the microscopic details become irrelevant and the critical properties of the 
system are dictated only by global and scale-invariant characteristics such 
as the dimensionality, symmetry and sometimes the topology. Because these 
most basic attributes are all discrete, field theories are readily classified 
and phase transitions can be categorized by their universality class.

One of the organizing principles of modern condensed matter is the concept 
of ``emergence,'' meaning large-scale patterns of behaviour that cannot be 
predicted from a knowledge of the short-range interactions. Quantum magnetic 
materials and models are widely recognized for the wealth of emergent 
phenomena they exhibit at low energies, which include multiple types of 
fractional excitation and of quantum spin liquid.\cite{Savary_RPP_2017}
Many more phenomena emerge when a system is driven into the critical 
regime around a phase transition.\cite{Sachdev_Book_2011} Beyond the 
continuous (second-order) transitions that are now well studied in 
experiment,\cite{Zapf_RMP_2014,Merchant_NP_2014} theory predicts that 
the order parameters of two phases with unrelated symmetries can vanish 
continuously and simultaneously. This deconfined quantum critical point 
(DQCP),\cite{Senthil_Science_2004,Fisher_PRB_2004} or multicritical 
point,\cite{Zhao_PRL_2020,Lu_PRB_2021} should be accompanied by 
emergent fractional excitations, exhibit unconventional critical 
scaling\cite{HuiShao_Science_2016} and possess an enhanced continuous 
symmetry. More generally, emergent enhanced symmetries have recently 
been discussed at a first-order transition\cite{Zhao_NP_2019} and at 
topological phase transitions.\cite{Zhu_PRL_2019,Schuler_SP_2023}

\begin{figure*}[t]
\centering
\includegraphics[width=\linewidth]{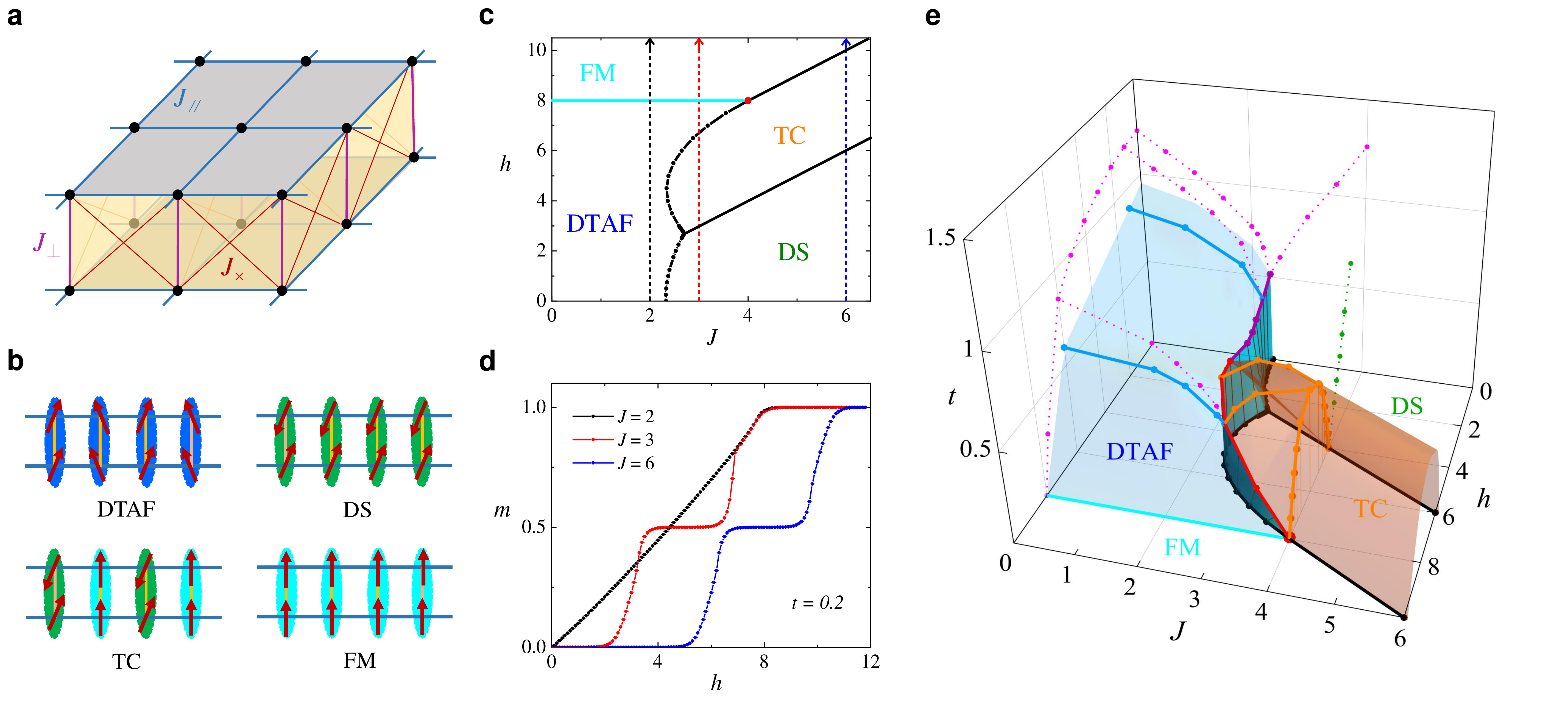}
\caption{{\bf Phase diagram of the fully frustrated bilayer Heisenberg model 
in an applied magnetic field.}
{\bf a} FFB model. Quantum spins ($S = 1/2$) are located at every site of a 
pair of square lattices. The interlayer dimer unit has magnetic interaction 
$J_{\bot}$, the intralayer interaction is $J_{\parallel}$ and the interlayer 
interaction between adjacent sites is $J_{\times}$; all three interactions 
are antiferromagnetic (AF) and of Heisenberg type.
{\bf b} Representations of the four different ground states in an applied 
field, the dimer triplet antiferromagnet (DTAF), dimer singlet (DS), 
checkerboard triplon crystal (TC) and fully polarized ferromagnet (FM).
{\bf c} Ground-state phase diagram obtained by the iPESS method. Apart from 
the field-driven DTAF-FM transition (cyan line), which is continuous, all 
other transitions are first-order (black lines). The red circle marks the 
quantum critical endpoint (QCEP) at $J = 4$ and $h = 8$.
{\bf d} Magnetization shown as a function of $J$ at low temperature 
($t = 0.2$), calculated by quantum Monte Carlo (QMC) for a system of 
size $L$$\times$$L$ dimers, with $L = 24$. The half-magnetization plateau 
characterizes the TC state.
{\bf e} Thermal phase diagram. Long-ranged DTAF and FM order is present only 
at zero temperature. The TC order parameter persists at finite temperatures, 
and this order melts continuously at the orange surface. The wall of 
discontinuities in the triplet density that separates the DTAF and DS phases 
terminates at a line of Ising critical points (purple) at finite temperatures, 
while the wall separating the DTAF and TC phases terminates at a line of 
emergent multicritical points (red). Each point on the red line is 
simultaneously the endpoint of a BKT transition of the DTAF (blue line with 
stars) and of a thermal Ising transition of the TC (orange), and exhibits an 
emergent 4-state Potts criticality. At high fields, the line of emergent 
multicritical points terminates at the QCEP (red circle). The magenta and 
green dashed lines mark characteristic crossover temperatures determined 
by computing the specific heat.}
\label{fig:1}
\end{figure*}

The many-body states of quantum spin systems can be altered by a variety of 
experimental methods, including an applied magnetic field,\cite{Zapf_RMP_2014} 
a hydrostatic pressure\cite{Merchant_NP_2014,Zayed_NP_2017} and controlled 
substitional disorder,\cite{Yu_Nat_2012} to obtain a wide range of 
possibilities for the investigation of phase transitions and related emergent 
phenomena. The field-induced magnetic order observed in dimerized spin systems 
can be described as a Bose-Einstein condensation of triplet excitations into 
the singlet ground state,\cite{Zapf_RMP_2014} while the pressure-induced 
transition\cite{Merchant_NP_2014} is a triplet condensation in the 3D XY 
universality class. It was pointed out recently that critical phenomena in 
quantum magnets are not restricted to second-order QPTs, but that the 
combination of quantum and thermal fluctuations can produce a critical point 
when the QPT is first-order.\cite{Wessel_PRL_2018} Using the $S = 1/2$ 
``fully frustrated bilayer'' (FFB) model shown in Fig.~\ref{fig:1}a, it was 
found that the discontinuity across the QPT, between the dimer-singlet and 
dimer-triplet phases (DS and DTAF in Fig.~\ref{fig:1}b), decreases with 
increasing temperature and terminates at a finite-temperature critical point. 
In this minimal model, there is no spontaneous symmetry-breaking across the 
line of discontinuities (meaning at $T > 0$, by the Mermin-Wagner theorem) 
and the extent of singlet-triplet order provides a quantum magnetic analogue 
of the liquid-gas transition, the two-component nature conferring an Ising 
universality.\cite{Chaikin_Book_1995}

This type of physics and its extensions have recently been pursued in a 
number of frustrated quantum spin models,\cite{Weber_SP_2022,Weber_PRB_2022,
Strecka_PRB_2023} but came to the fore when it was shown to be the origin of 
critical-point behaviour found\cite{Mila_Nature_2021} in specific-heat 
measurements on the frustrated quantum antiferromagnet SrCu$_2$(BO$_3$)$_2$. 
This compound provides a remarkably faithful realization of the 
Shastry-Sutherland model (SSM),\cite{Shastry_PBC_1981} not only at 
ambient pressure\cite{Kageyama_PRL_1999} but also in its pressure-induced 
QPTs.\cite{Zayed_NP_2017} However, the first-order QPT in SrCu$_2$(BO$_3$)$_2$ 
and the SSM separates a DS phase from a plaquette-singlet phase, which does 
have long-ranged order at low temperatures and a continuous thermal 
transition,\cite{Mila_Nature_2021} making the situation more complex than 
the FFB. SrCu$_2$(BO$_3$)$_2$ also shows a complex cascade of QPTs in an 
applied magnetic field,\cite{Kageyama_PRL_1999,Matsuda_PRL_2013} raising 
important questions about the nature of criticality under combined fields and 
pressures. The possibilities range from emergent enhanced symmetries and 
emergent types of multicriticality to the appearance of a DQCP suggested 
by theory\cite{Lee_PRX_2019} and experiment.\cite{Cui_Science_2023}

\begin{figure*}[t]
\includegraphics[width=0.96\linewidth]{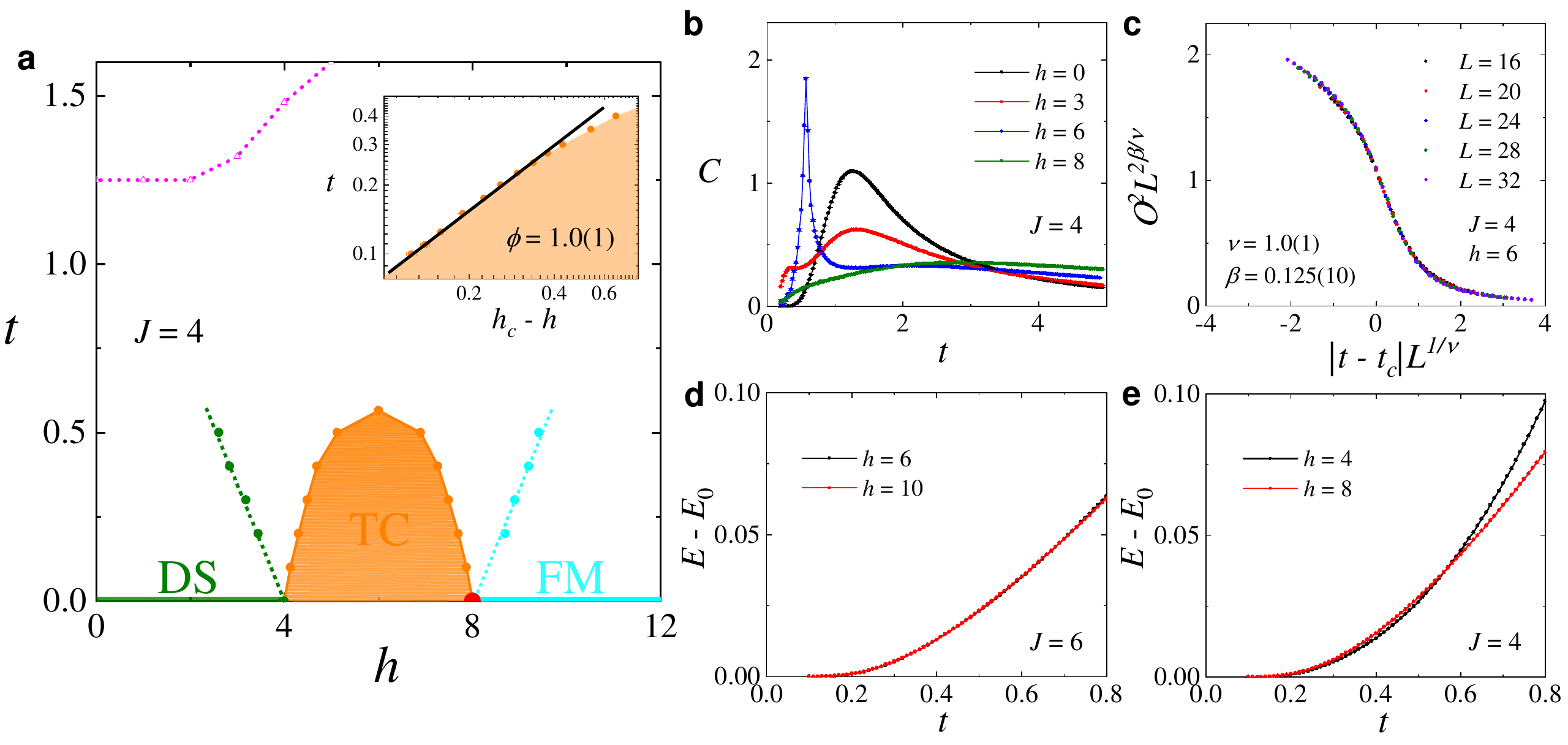}
\caption{{\bf Physics of the DS and TC phases, and at the QCEP.}
{\bf a} ($h, t$) phase diagram of the FFB model at $J = 4.0$. The dashed 
green line shows the gap of the DS phase, the dashed cyan line the FM gap, 
the dashed magenta line the characteristic crossover temperature determined 
from the specific heat and the solid orange line the TC ordering temperature. 
The inset shows that the field-scaling exponent is $\phi = 1$; the model does 
not have 2+1D Ising universality because the $t = 0$ transition is first-order. 
{\bf b} Specific heat, $C(t)$, shown for a number of applied field values. The 
peak temperatures are used to determine the phase-boundary (orange) and 
crossover (magenta) lines in panel a. 
{\bf c} Ising nature of the thermal transition of the TC phase, shown by 
scaling collapse of the order parameter, $O (t,L)$, calculated at $J = 4$ 
and $h = 6$ for systems of linear size $L$. The critical exponents are 
fully compatible with the Ising values $\nu = 1$ and $\beta = 1/8$.
{\bf d} Thermal energy, $E(t) - E_0$, computed with $L = 24$ and shown at the 
points $(J, h) = (6, 6)$ and $(J, h) = (6, 10)$, which lie respectively on the 
zero-temperature first-order DS-TC and TC-FM lines. The exponential form arises 
due to the gapped nature of both phases and appears identical as a consequence 
of particle-hole symmetry about triplet density $n_t = 1/2$.
{\bf e} $E(t) - E_0$ at the points $(J, h) = (4, 4)$, on the DS-TC 
zero-temperature first-order line, and $(J, h) = (4, 8)$, which is the QCEP. 
The change in exponential form indicates a violation of particle-hole symmetry 
due to the presence of low-energy spin-wave excitations at the QCEP.}
\label{fig:2}
\end{figure*}

Here we look more deeply into the FFB model to study 
how its phases and phase transitions evolve in an applied magnetic field. 
The field enriches the phase diagram, turning the dimer-triplet state into 
a Berezinskii-Kosterlitz-Thouless (BKT) phase with quasi-long-ranged order 
(qLRO), and the DS state into a checkerboard triplon crystal (TC) phase with 
LRO at finite temperatures. The Ising critical point becomes a critical 
line, first retaining its Ising character but then gaining an emergent 4-state 
Potts symmetry on the multicritical phase boundary to the TC regime, before 
terminating at a quantum critical endpoint (QCEP). All our thermal calculations 
are performed using large-scale quantum Monte Carlo (QMC) methods enabled by 
the recent qualitative breakthrough that the fully frustrated system formulated 
in the dimer basis has no sign problem. We map the quantum FFB model to a 
classical equivalent and then to the Ashkin-Teller model in order to trace 
the origin of emergent criticality, and hence to shed light on its possible 
appearance in highly frustrated quantum magnetic materials such as 
SrCu$_2$(BO$_3$)$_2$. 

\begin{figure*}[t]
\includegraphics[width=0.96\linewidth]{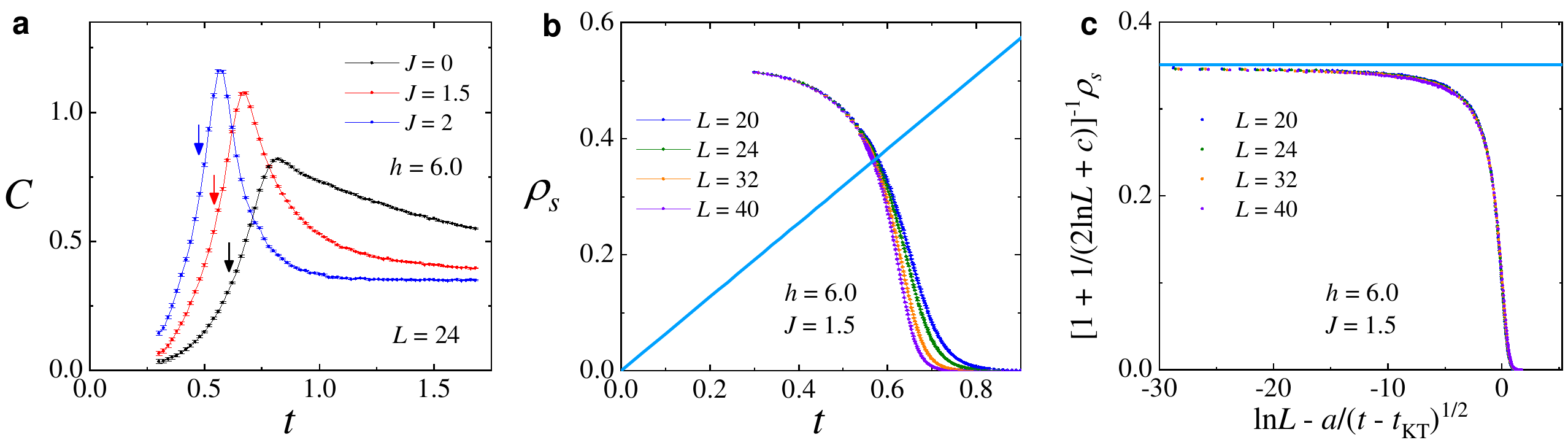}
\caption{{\bf Physics of the DTAF phase.} 
{\bf a} Thermodynamic properties of the DTAF, illustrated by the specific 
heat, $C(T)$, at $h = 6$ for three different $J$ values. The centre of the 
single broad peak gives a characteristic crossover temperature that lies 
above the BKT transition (coloured arrows).  
{\bf b} Spin stiffness, $\rho_s(t)$, computed for systems of different 
sizes, $L$, at $h = 6$ and $J = 1.5$. The BKT transition temperature, 
$t_{\rm KT}$, is determined from the crossing point of $\rho_s (t)$ with the 
linear function $2t/\pi$ (cyan line) in the limit $L \rightarrow \infty$.
{\bf c} Finite-size scaling of the spin stiffness near the BKT transition, 
illustrated by its collapse according to the form $\rho_s(t,L) = [1 + 1/(2 
\ln L + c)] F_{\rho} (\ln L - a/\sqrt{t - t_{\rm KT}})$, where $F_{\rho}$ is a 
scaling function. We extract the parameters $a = 1.1 \pm 0.05$, $c = 0.05 
\pm 0.02$ and $t_{\rm KT} = 0.549 \pm 0.002$.}
\label{fig:3}
\end{figure*}

\medskip
\noindent
{\large {\bf Results}}

The Hamiltonian of the FFB model is 
\begin{eqnarray}
\mathcal{H} \, = \, & & \sum_i {J_{\bot} \vec{S}_{i,1} \cdot \vec{S}_{i,2}}
 - H \sum_{i, m = 1, 2} {S_{i,m}^{z}} \\ & & + \sum_{< i,j >, m = 1, 2} 
{\left[ J_{\parallel} \vec{S}_{i,m} \cdot \vec{S}_{j,m} + J_{\times} 
\vec{S}_{i,m} \cdot \vec{S}_{j,\bar{m}} \right]}, \nonumber 
\end{eqnarray}
where $\vec{S}_{i,m}$ is a quantum $S = 1/2$ spin at site $i$ and layer $m$ 
of the square-lattice bilayer shown in Fig.~\ref{fig:1}a, $H$ is the applied 
magnetic field and the antiferromagnetic (AF) Heisenberg interactions are 
$J_{\bot}$ on the interlayer dimer bond, $J_{\parallel}$ within each square 
lattice and $J_{\times}$ between next-neighbour interlayer sites, which 
frustrates $J_{\parallel}$. We consider only the fully frustrated case, $J_{\times}
 = J_{\parallel}$, where the model can be rewritten in the dimer basis as
\begin{equation}
\label{Eq:HamTriplet}
\mathcal{H} = J_{\parallel} \sum_{i,j}{\vec{T}_i \cdot \vec{T}_j}
 + J_{\bot} \sum_i ({\textstyle \frac12} \vec{T}_{i}^{2} - 
{\textstyle \frac{3}{4}}) - H \sum_i {T_{i}^{z}},
\end{equation}
with $\vec{T}_i = \vec{S}_{i,1} + \vec{S}_{i,2}$ the total spin of each dimer; 
$\vec{T}_i^2$ is proportional to the spin-triplet density and is locally 
conserved, meaning on every dimer, $i$. This is the property that causes the 
sign problem, which conventionally accompanies QMC simulations on frustrated 
spin systems, to be completely absent.\cite{Alet_PRL_2016,Bruce_PRB_2016} 
Thus we can use the stochastic series expansion (SSE) 
algorithm\cite{Syljuasen_PRE_2002} to obtain highly accurate simulation 
results for square-lattice dimensions $L$$\times$$L$ up to a linear size 
of $L = 40$. We take $J_{\parallel}$ as the unit of energy and define the 
reduced coupling $J = J_{\bot}/J_{\parallel}$, reduced field $h = H/J_{\parallel}$ 
and reduced temperature $t = T/J_{\parallel}$, with the lowest temperature we 
access being $t = 0.1$. To complement these thermal results, we calculate 
the quantum ($t = 0$) phase diagram of the FFB in a field by applying the 
tensor-network method of infinite Projected Entangled Simplex States 
(iPESS),\cite{Xie_PRX_2014} as summarized in the Methods section. The FFB model 
has been studied in detail by SSE QMC at zero field,\cite{Wessel_PRL_2018} and 
here we reveal the complex and emergent phenomena induced by the magnetic 
field. Some properties of the FFB in a field have been investigated by 
Richter and coworkers,\cite{Richter_PRB_2006,Oleg_PRB_2010,Richter_PRB_2018} 
although these authors did not discuss the full phase diagram or emergent 
critical properties. 

We begin by using iPESS to identify the four ground states of the model 
illustrated schematically in Fig.~\ref{fig:1}b. A straightforward energy 
comparison, described in Sec.~S1 of the Supplementary Information 
(SI),\cite{si} yields the $t = 0$ phase diagram shown in Fig.~\ref{fig:1}c. 
The DS and dimer-triplet antiferromagnet (DTAF) states are familiar at zero 
field. Beyond a finite field, the DS is driven into the intermediate TC 
state, of alternating dimer singlets and field-aligned triplets, and the 
magnetization shows a plateau at half of its saturation value 
(Fig.~\ref{fig:1}d).\cite{Oleg_PRB_2010} Sufficiently strong fields cause 
a full polarization of the DTAF and TC phases into the FM state. Because 
only the DTAF and FM phases have the same triplet density ($n_t = 1$), 
every phase boundary in Fig.~\ref{fig:1}c is first-order, other than 
the line of continuous DTAF-FM transitions. This line terminates at a
QCEP at $J = 4$ and $h = 8$, where it meets the first-order phase 
boundary of the TC state. 

Figure \ref{fig:1}e shows the thermal phase diagram we obtain from our SSE 
QMC simulations. Only the TC phase has a LRO, which is terminated at a 
thermal transition (orange surface), although the DTAF has qLRO that is 
lost at a BKT transition (blue). The first-order DS-TC and TC-FM transitions 
become continuous at any finite temperature, an unconventional property we 
give the simple terminology ``zero-temperature first-order line.'' The 
first-order DTAF-DS and DTAF-TC lines persist to finite temperatures, forming 
a wall of discontinuities across the phase diagram, which is terminated by a 
(multi)critical line whose nature is the primary focus of our work. 

To analyse the rich variety of phenomena on display in Fig.~\ref{fig:1}e, 
we start on the DS side by considering the field-induced behaviour at fixed 
$J = 4$ (Fig.~\ref{fig:2}a). The DS phase has no order parameter, but can be 
characterized by the dimer spin correlation.\cite{Mila_Nature_2021} At finite 
fields, the average of the three Zeeman-split triplon branches determines the 
position of the broad maximum in the specific heat (Fig.~\ref{fig:2}b), while 
the closure of the gap to the lowest triplon sets the DS-TC transition, shown 
in Fig.~\ref{fig:1}c. At $J = 4$ and $t = 0$, the ground state undergoes 
first-order transitions from DS to TC at $h = 4$ and TC to FM at $h = 8$, 
where the TC phase supports true LRO and thus has a continuous thermal phase 
transition at any point on the orange surface in Fig.~\ref{fig:1}e. By the 
scaling-collapse analysis presented in Fig.~\ref{fig:2}c and described in 
the Methods section, we show that this transition has the critical exponents 
of Ising universality, $\nu = 1$ and $\beta = 1/8$, as might be anticipated 
from the twofold degeneracy (i.e.~broken Z$_2$ sublattice symmetry) of the 
checkerboard TC state.  

\begin{figure*}[t]
\includegraphics[width=0.96\linewidth]{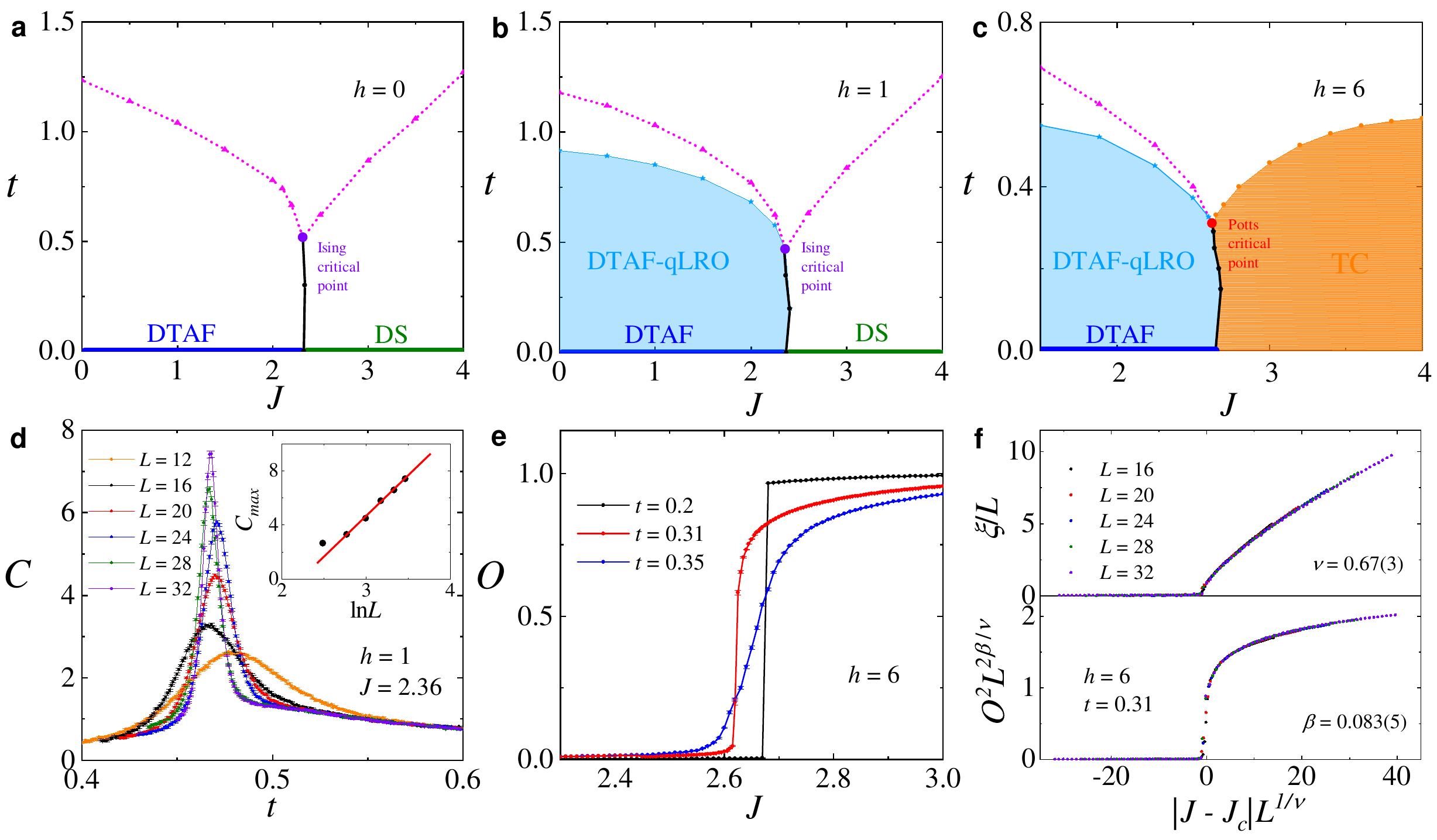}
\caption{{\bf Physics of the (multi)critical point.}
{\bf a}-{\bf c} Phase diagrams of the FFB model in the $(J,t)$ plane at 
$h = 0$ (a), $h = 1$ (b) and $h = 6$ (c). Black lines in each panel 
mark the first-order transition determined by the discontinuity in the 
triplet density. Magenta curves show crossover temperatures determined 
from the specific-heat peaks at each value of $h$ and $J$. Blue lines and 
stars mark the BKT transition of the DTAF, below which the system exhibits 
qLRO. Orange lines and circles show the continuous thermal transition of 
the TC. At each field, all of the transition and crossover lines meet at 
a critical point that terminates the line of first-order transitions: this 
point is located at $(J_{c}, t_{c}) = (2.315(1), 0.517(3))$ for $h = 0$, 
\cite{Wessel_PRL_2018} $(2.36(1), 0.47(1))$ for $h = 1$ and $(2.624(4), 
0.308(5))$ for $h = 6$, and has either Ising (purple circle) or 4-state 
Potts universality (red circle).
{\bf d} Specific heat, $C(t)$, calculated at $J = J_c$ for $h = 1$ using 
a range of system sizes, $L$. The sharpening of the peak with increasing 
$L$ signals a continuous transition and the scaling of the peak value, 
$C_{\rm{max}}$, with $\ln L$ (inset) indicates that the transition in the 
low-field regime is in the same 2D Ising universality class as the $h = 0$ case.
{\bf e} Order parameter, $O$, of the TC phase calculated with $L = 24$ for 
$h = 6$ and shown as a function of $J$ near the emergent (multi)critical 
point for three temperatures below, at and above $t_c$.
{\bf f} Scaling collapse of the finite-size correlation length, $\xi$, and
the order parameter, $O$, across the emergent (multi)critical point. The 
critical exponents estimated from these two types of data collapse are 
$\nu = 0.67 \pm 0.03$ and $\beta = 0.083 \pm 0.005$, which are fully 
consistent with 4-state Potts universality.}
\label{fig:4}
\end{figure*}

We reiterate the curious nature of the critical surface of the TC phase 
at the DS and FM transitions, which changes from second- to first-order 
precisely at $t = 0$. The termination of a second-order line on a first-order 
one is known as a critical endpoint,\cite{Fisher_PRL_1990} and a second-order 
line turning first-order is a tricritical point, but the DS-TC and TC-FM 
boundaries lack a first-order surface ({\it cf}.~the QCEP at ($J = 4$, $h
 = 8$)) or half-surface; thus we use instead the term zero-temperature 
first-order line. The physics of the DS-TC line is that any state excluding 
nearest-neighbour triplon pairs minimizes the energy, resulting in a highly 
degenerate ground state.\cite{Richter_PRB_2006, Oleg_PRB_2010, 
Richter_PRB_2018} Away from the line, these states form low-energy excitations, 
while excitations containing at least one triplon pair have a gap of order 
$J_{\parallel}$, and these two types of process account for the two peaks in 
$C(t)$ (Fig.~\ref{fig:2}b). Exactly analogous behaviour is observed around 
the TC-FM line as a consequence of particle-hole symmetry about $n_t = 1/2$. 
By contrast, the TC-DTAF transition remains first-order up to a finite 
temperature, where the TC surface meets the line of critical points to 
establish the emergent criticality we analyse below.  

Finally, to investigate the QCEP, defined as the termination point of a 
line of continuous QPTs (the DTAF-FM line), we characterize the low-energy 
excitation spectrum by computing the thermal energy, $E(t) - E_0$, at different 
points on the zero-temperature first-order lines. As Fig.~\ref{fig:2}d shows 
for the two transitions at $J = 6$, $E(t)$ has the same exponential form, 
indicating gapped excitations above the ground manifold, with the same 
prefactor, thereby respecting the particle-hole symmetry. However, in 
Fig.~\ref{fig:2}e we observe a departure from this symmetry when the QCEP is 
compared with its counterpart ($h = 4$), with additional thermal energy at 
$t \lesssim 0.5$ suggesting low-energy AF spin-wave fluctuations, in 
finite-sized regimes of the neighbouring DTAF phase, appearing within the gap. 

\begin{figure*}[t]
\includegraphics[width=0.96\textwidth]{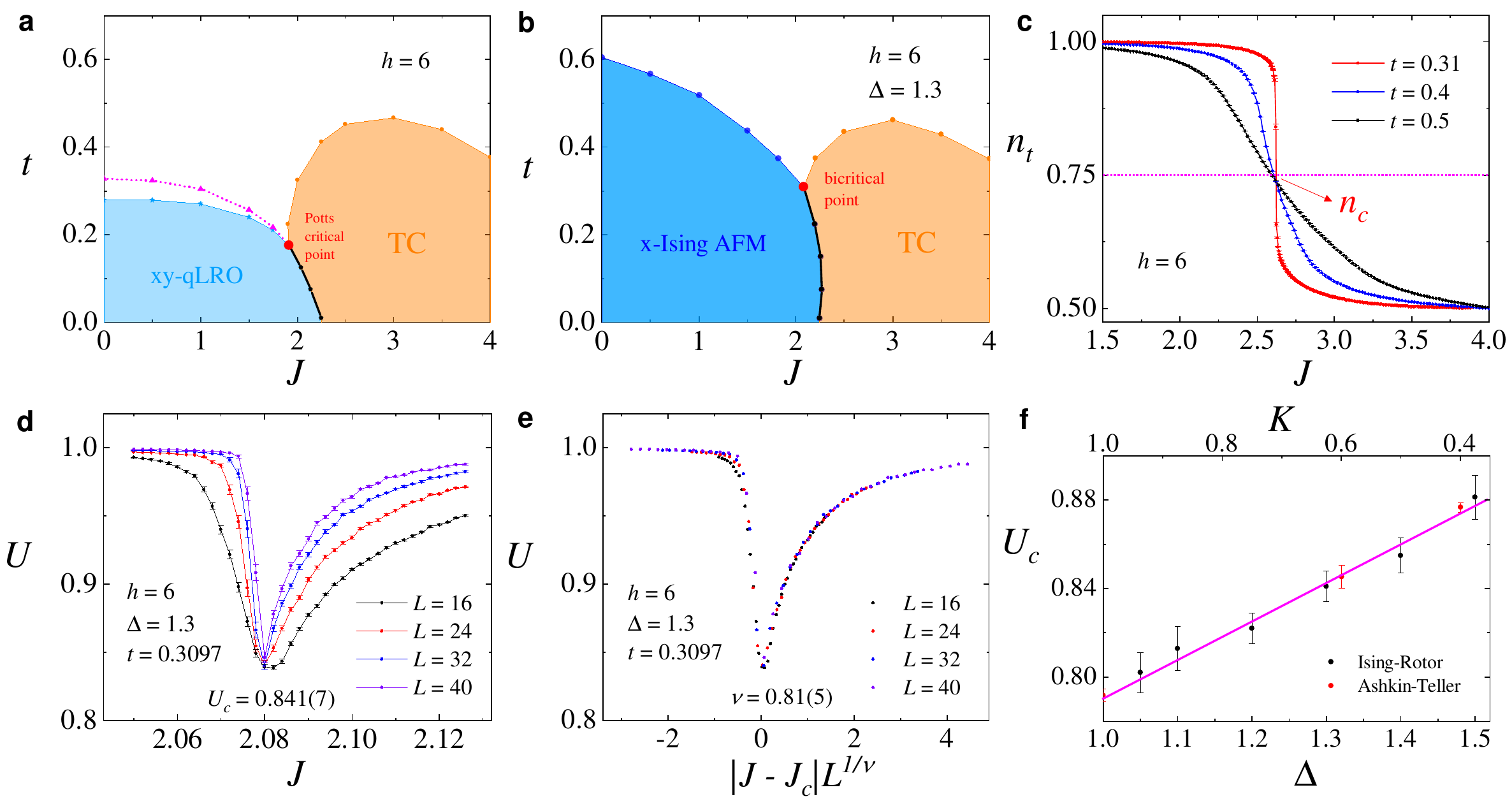}
\caption{{\bf Classical model, spin-anisotropic model and Ashkin-Teller 
model.}
{\bf a} Phase diagram of the classical spin-rotor model extracted from 
the FFB model, shown at $h = 6$.
{\bf b} Phase diagram of the classical model with spin anisotropy, shown 
at $h = 6$ with $\Delta = 1.3$ (weak Ising anisotropy). The qLRO of the 
DTAF phase becomes a true LRO with a thermal Ising transition occurring 
at a rather higher ordering temperature. Because the TC phase retains its 
Ising transition, the emergent critical point becomes a bicritical point 
(red circle) where the two Ising transition lines meet at $(J_{c}, t_{c})
 = (2.0792(14), 0.3097(8))$.
{\bf c} Triplet density, $n_t$, computed for $L = 24$ and shown as a 
function of $J$ for various temperatures at $h = 6$. $n_t = 0.75$ at the 
multicritical point and reflects the approximate particle-hole symmetry 
between the DTAF and TC phases at their transition. 
{\bf d} Binder cumulant, $U$, computed as a function of $J$ across the 
bicritical point with $h = 6$ and $\Delta = 1.3$ for a range of system 
sizes. $J_c = 2.0792 \pm 0.0014$ is determined from the touching point of 
the Binder-cumulant curves. 
{\bf e} Collapse of the Binder-cumulant data shown in panel d, from which 
we estimate the non-universal critical exponent $\nu = 0.81 \pm 0.05$.
{\bf f} Critical value of the Binder cumulant at the bicritical point, $U_c$ 
(black circles), shown as a function of the spin anisotropy. For comparison 
we show $U_c$ values at the critical points of the Ashkin-Teller model (ATM, 
red circles). The magenta line is a linear fit which shows that extrapolating 
$U_c$ to the spin-isotropic limit ($\Delta = 1$) gives good agreement with 
the critical value obtained at $K/J = 1$ in the ATM, where this model has 
4-state Potts universality.}
\label{fig:5}
\end{figure*}

Turning to the DTAF side of the phase diagram (Fig.~\ref{fig:1}e), the field 
breaks the SU(2) spin symmetry down to U(1) and the DTAF phase supports 
qLRO below a finite-temperature BKT transition.\cite{Berezinskii_SPJ_1972,
Kosterlitz_JPC_1972} In Fig.~\ref{fig:3}a we use the specific heat to show 
that the thermodynamic properties of the DTAF phase, computed at $h = 6$ for 
a number of $J$ values, remain similar to those at $h = 0$, with a single 
peak marking a characteristic crossover temperature. An accurate determination 
of $t_{\rm KT}$ can be obtained\cite{Weber_PRB_1988,Sandvik_Book_2010} from the 
finite-size scaling of the spin stiffness, $\rho_s(t)$, as we explain in the 
Methods section and show in Figs.~\ref{fig:3}b and \ref{fig:3}c. As 
expected,\cite{Chaikin_Book_1995} $t_{\rm KT}$ is not reflected in the 
conventional thermodynamic response (Fig.~\ref{fig:3}a), lying systematically 
beneath the crossover such that the two temperatures form two sets of surfaces 
that meet along the line of (multi)critical points (Fig.~\ref{fig:1}e); in 
Sec.~S2 of the SI we show further data illustrating this situation in the 
DTAF phase. 

Given the qLRO of the DTAF phase below $t_{\rm{KT}}$, and the LRO of the 
TC phase below a transition that also appears to converge on the critical 
line, the crucial question arises of how these field-induced phases affect 
the universality. We preface the discussion to follow with an important 
remark: we have found in Figs.~\ref{fig:4}a-c that all the transition and 
crossover lines meet with an accuracy of 0.01 in our units, as we show in 
Sec.~S2 of the SI.\cite{si} Rather than perform extensive numerical 
calculations to achieve a higher accuracy, which still would not serve as 
a proof of exact convergence, we base our discussions, particularly of the 
multicritical DTAF-TC point, on the analysis of additional symmetries and 
related models below. At zero field (Fig.~\ref{fig:4}a), where neither the 
DS nor the DTAF possesses finite-$t$ order due to symmetry-breaking, using 
$n_t$ as an effective order parameter reveals a liquid-gas-type 
transition,\cite{Wessel_PRL_2018} where the first-order line is terminated 
by a critical point with Ising universality, and crossover lines (determined 
from the specific-heat peaks, Sec.~S2 of the SI\cite{si}) appear on both 
sides of the transition.\cite{Weber_SP_2022}

At a fixed low field, where the DTAF has a BKT transition that meets the 
first-order line at the critical point (Fig.~\ref{fig:4}b), we investigate 
the nature of criticality by computing the specific heat as a function of 
system size, as shown in Fig.~\ref{fig:4}d. The progressive sharpening of 
the peak can be characterized by its height, $C_{\rm max} (L)$, which in the 
inset we find follows precisely the $\ln L$ scaling of Ising 
universality.\cite{Wessel_PRL_2018} Thus we conclude that the BKT 
transition has no effect on the universality of the critical point 
in this instance, and we explain this result below. 

Turning now to the DTAF-TC transition, in Fig.~\ref{fig:4}c we show the 
situation at $h = 6$ and in Sec.~S2 of the SI\cite{si} we present results 
elsewhere on this phase boundary. Figure \ref{fig:4}e shows explicitly 
how the discontinuity in the TC order parameter at $t < t_c$ becomes smooth at 
and above $t_c = 0.31$. At minimum (neglecting the BKT transition), the merging 
of the discontinuous DTAF-TC line with the continuous thermal transition of the 
TC phase should change the Ising critical point to a tricritical Ising point. 
To determine how the universality is altered, in Fig.~\ref{fig:4}f we consider 
the scaling collapse of all our finite-$L$ data at $t_c$ for $h = 6$ and 
extract the critical exponents $\nu = 0.67 \pm 0.03$ and $\beta = 0.083 
\pm 0.005$. These values are far from the Ising and tricritical Ising 
cases, instead corresponding very well to 4-state Potts universality, 
where $\nu = 2/3$ and $\beta = 1/12$.\cite{Potts_RMP_1982}

{\it {A priori}}, this Potts criticality comes as a complete surprise, because 
the FFB model in a field has no S$_4$ permutation symmetry, and thus it 
satisfies all the criteria of an emergent phenomenon. Our discovery of such an 
exact 4-state Potts universality also appears highly unlikely if the meeting of 
the three transition lines at a single multicritical point were not exact. To 
gain further insight, we follow a series of mappings to unveil the underlying 
symmetries of the system, and summarize the procedure here (a more complete 
discussion is presented in Sec.~S3 of the SI\cite{si}). First we map the 
quantum model of Eq.~\eqref{Eq:HamTriplet} to a classical model consisting of 
an O(3) rotor, representing the spin degrees of freedom, coupled to an Ising 
variable representing the triplon density. It is easy to perform large-scale 
Monte Carlo simulations of this model, and in Fig.~\ref{fig:5}a we show that 
the phase diagram at $h = 6$ is very similar, even semi-quantitatively, to the 
quantum one (Fig.~\ref{fig:4}c). A BKT transition arises for the rotors in 
an applied field and a density-driven Ising transition breaks the sublattice 
symmetry. Across the critical point where the BKT and Ising transitions meet, 
we again find $\nu \simeq 2/3$ and $\beta \simeq 1/12$, meaning that the 
classical model also has 4-state Potts universality. 

Next we introduce an Ising-type spin anisotropy, $\Delta > 1$, that breaks 
the U(1) spin symmetry to Z$_2$, turning the BKT transition into a thermal 
Ising transition, below which the system has true LRO that we denote as 
``x-Ising'' in Fig.~\ref{fig:5}b. In this situation, the two Ising transitions 
meet at a bicritical point, below which the transition directly from x-Ising 
to TC is first-order. The bicritical scenario provides a well accepted instance 
where the three transition lines do meet at a single point, as our numerical 
results indicate for the FFB model (Fig.~\ref{fig:4}c). The correlation-length 
exponent, $\nu$, that we obtain from scaling collapse at the bicritical point 
varies between $2/3$ and $1$ with the strength of spin anisotropy, as we show 
in Sec.~S3 of the SI \cite{si}, and thus the anisotropic classical model has 
non-universal behaviour.

Because both the DTAF and TC phases break the sublattice symmetry, each 
has two degenerate configurations,\cite{footnote1} and in total four 
ground-state configurations are degenerate at the transition. Here the 
triplon density drops from $n_t = 1$ in the DTAF to $n_t = 1/2$ in the TC, 
and the evolution of $n_t$ with $J$ across the emerging critical point 
indicates a particle-hole symmetry about $n_t = 3/4$ (Fig.~\ref{fig:5}c). 
Defining the Ising variables $\sigma_i$, associated with the sublattice 
symmetry, and $\eta_i$, associated with the particle-hole symmetry, the Ising 
variable $\tau_i = \sigma_i \eta_i$ obeys the relations $\tau \rightarrow - 
\sigma$ and $\sigma \rightarrow - \tau$ under spin inversion. With these 
variables we construct the minimal effective model that should describe 
the critical properties around the emerging critical point in the form
\begin{equation}
H_{\rm eff} = - J \sum_{i,j} (\sigma_i \sigma_j + \tau_i \tau_j) - K \sum_{i,j} 
(\sigma_i \sigma_j)(\tau_i \tau_j),
\end{equation}
as discussed in detail in Sec.~S3 of the SI.\cite{si} This model is precisely 
the Ashkin-Teller model (ATM). It is well known\cite{Wiseman_PRE_1993} that 
the ATM exhibits non-universal exponents, which lie between Ising and Potts 
universalities, along the critical line obtained for $0 \leq K/J \leq 1$.

To analyse the correspondence between this ATM and the classical model with 
variable spin anisotropy, we define a composite order parameter, $O_b^2 = 
O_{\rm x}^2 + O_{\rm TC}^2$, that contains the order of both the x-Ising and TC 
phases (Sec.~S3C of the SI\cite{si}). We then consider the Binder cumulant 
associated with $O_b$, $U = 2(1 - \langle O_b^4 \rangle / 2 \langle O_b^2 
\rangle^2)$, which has the property that $U \rightarrow 1$ deep in both 
ordered phases, but dips across the bicritical point. Figure \ref{fig:5}d 
shows this evolution of $U$ for a range of system sizes, from whose touching 
point we determine both the location of the bicritical point and the value 
$U_c$. Because of its dimensionless nature, $U_c$ should also reflect the 
universality of the critical point,\cite{Goldenfeld_Book_1992,Dohm_PRE_2004,
Selke_JPA_2005,Songbo_PRL_2012} and the scaling collapse shown in 
Fig.~\ref{fig:5}e presents an accurate means of extracting $\nu$. For the 
spin-anisotropic model, we find that both $U_c$ and $\nu$ vary with the 
anisotropy $\Delta$, exhibiting a non-universal evolution such that $\nu$ 
at the bicritical point varies between $1$ and $2/3$, signalling a continuous 
change between Ising and 4-state Potts universality. The monotonic dependence 
of $U_c$ on $\Delta$, shown in Fig.~\ref{fig:5}f, compares exactly with the 
$K/J$ scaling of the ATM, and its extrapolated value as $\Delta \rightarrow 1$ 
agrees well with the value $U_c \rightarrow 0.792 \pm 0.003$ found for the ATM 
at the Potts point.\cite{Songbo_PRL_2012}

\medskip
\noindent
{\large {\bf Discussion}}

We have presented the global phase diagram of the FFB Heisenberg model in 
an applied magnetic field, showing how the field allows systematic control 
over the phase competition. We have explained the rich variety of phase 
transitions and (multi)critical lines or points, and found numerical evidence 
for a striking example of emerging critical behaviour. This emergent 4-state 
Potts universality, arising along the line of multicritical points where the 
BKT transition of the DTAF and the Ising one of the TC phase meet, implies 
that the multicritical system has a higher S$_4$ permutation symmetry that 
the microscopic spin model does not possess. This result is clearly different
from previous studies of transitions between BKT and Ising phases, such as 
the well known FFXY [U(1)$\otimes$Z$_2$] model,\cite{Song_PRB_2022} of 
anisotropic Heisenberg spin models\cite{Holtschneider_PRB_2007} and of 
higher symmetries emerging from combined Ising order 
parameters.\cite{Zhu_PRL_2019,Schuler_SP_2023} In the FFB, a symmetry 
analysis reveals an additional particle-hole symmetry that lies beyond all of 
these models and can be used to map the critical system to an effective ATM. 
The generic ATM has a wide parameter regime where one line of critical points 
has nonuniversal and continuously varying exponents, as we find in the model 
with spin anisotropy. In the isotropic limit, where the spin symmetry is 
enhanced to U(1), we expect 4-state Potts universality (i.e.~an enhanced S$_4$ 
symmetry), because the Potts point is the only multicritical point with 
enhanced symmetry in the ATM.\cite{Dijkgraaf_CMP_1988} Precisely how this 
effective S$_4$ symmetry emerges in the critical behaviour as the spin 
symmetry is restored to U(1) is a well defined problem that deserves 
further exploration in the expanding field of emergent phenomena.

We found at low fields that the BKT phase of the DTAF has no effect on the 
Ising nature of the zero-field critical point. This result is explained by 
the fact that the phase mode associated with the BKT transition couples only 
marginally to the Ising variable (the triplet density, which is an amplitude 
mode), as we discuss in Sec.~S3 of the SI.\cite{si} At fields high enough to 
create the TC phase, the BKT nature of the DTAF again plays no role in 
determining the properties, including the emergent Potts universality, of the 
multicritical line: our construction of the ATM relied on the symmetries of 
the system, but not on the presence of qLRO when the spin anisotropy is 
removed. The origin of this result may be traced by considering the 
conformally invariant field theories (CFTs) describing the criticality in 
two-dimensional systems.\cite{Friedan_PRL_1984} The ATM with $K \le J$ is 
described by a $c = 1$ CFT. The BKT transition also has $c = 1$, and hence 
its action on the ATM is only marginal, with the result that the universality 
class is not altered. 

We expect that the multicritical points, emergent critical properties 
and QCEP we find in the FFB model are directly relevant to a number of 
dimerized and frustrated quantum magnetic materials. The magnetic properties 
of Ba$_2$CoSi$_2$O$_6$Cl$_2$ were suggested to be those of the FFB model with 
an in-plane spin anisotropy,\cite{Tanaka_JPSJ_2014} and one might anticipate 
exporing some of the FFB phase diagram under combined field and pressure. 
The zero-field Ising critical point of the FFB has been found in 
SrCu$_2$(BO$_3$)$_2$,\cite{Mila_Nature_2021} and our results for the TC 
phase boundaries should help to interpret the transitions to the plaquette 
phase of this system (both have broken Z$_2$ symmetry). SrCu$_2$(BO$_3$)$_2$ 
exhibits a cascade of fractional magnetization plateau states at higher 
applied fields,\cite{Matsuda_PRL_2013} and at higher pressures has the 
plaquette and AF phases of the SSM,\cite{Zayed_NP_2017} raising the prospect 
of multiple opportunities to search for emergent critical behaviour along 
the associated phase-transition lines.\cite{Shi_NC_2022} Thus we expect our 
results to provide a useful foundation for new theoretical and experimental 
studies of criticality and emergent phenomena.

\medskip
\noindent
{\large {\bf Methods}}

We study the quantum phase diagram of the FFB model using the PESS 
tensor-network method.\cite{Xie_PRX_2014} The simplex basis allows a highly 
efficient encoding of the entanglement in frustrated quantum spin systems, 
and its construction is described in Sec.~S1 of the SI.\cite{si} The lattice 
translation symmetry is used to work on a spatially infinite system and the 
ground state is found by evolution in imaginary time, for which a 
simple-update method is sufficient in the FFB. Because three of the four 
ground states are known exactly, detailed calculations pushing the limits 
of the trucation parameter ($D$, the tensor bond dimension) are not required 
to achieve accurate convergence (Sec.~S1).  

As noted above, the total absence of a sign problem in the fully frustrated 
model ($J_{\parallel} = J_{\times}$) makes SSE QMC simulations possible on large 
systems and at low temperatures. To extract the maximum accuracy from our 
simulations, in particular for the determination of critical points and 
exponents, we exploit the finite-size scaling properties of the known phases. 

The transition to the TC phase is studied by defining the dimer-triplet 
correlation function
\begin{equation}
\label{Eq:TripletCorr}
R (\vec{k}) = L^{-2} \sum_{ij} {e^{i \vec{k} \cdot (\vec{r}_i - \vec{r}_j)} 
\left\langle (T_{i}^{2} - 1) (T_{j}^{2} - 1) \right\rangle},
\end{equation}
from which we obtain the correlation length 
\begin{equation}
\label{Eq:xi}
\xi = \frac{2\pi}{L} \sqrt{R (\vec{Q})/R (\vec{Q} + \delta\vec{Q}) - 1}
\end{equation}
and the order parameter 
\begin{equation}
\label{Eq:O}
O^2 = R(\vec{Q})/L^2,
\end{equation}
where $\vec{Q} = (\pi,\pi)$ is the ordering wavevector and $\delta\vec{Q}
 = (2\pi/L,0)$. We obtain the critical exponents by establishing the 
scaling collapse of both quantities in the forms\cite{Sandvik_Book_2010}
\begin{eqnarray}
 \xi & = & L F_{\xi,g} \big( |g| L^{1/\nu} \big), \\
 O^2 & = & L^{2\beta/\nu} F_{O,g} \big( |g| L^{1/\nu} \big),
\end{eqnarray}
where $|g|$ denotes the three quantities $|J - J_c|$, $|h - h_c|$ and 
$|t - t_c|$ that we use to effect a systematic approach to the critical 
points when the other variables are fixed to their critical values. $F$ 
denotes a single function that describes all the data when scaled in this 
way, allowing a highly accurate determination of the critical exponents. 

To study the BKT transition, we calculate the spin stiffness from the 
expression $\rho_s = {\textstyle \frac12} t \langle w_{x}^{2} + w_{y}^{2} 
\rangle$, in which $w_{\alpha} = \sum_b ( N_{b,\alpha}^+ - N_{b,\alpha}^-)/L$ 
with $\alpha$ denoting the direction $x$ or $y$ and $N_{b,\alpha}^{\pm}$ 
expressing the number of operators $T_{i(b)}^{\pm} T_{j(b)}^{\mp}$ associated 
with bond $b$ in direction $\alpha$ appearing in the QMC operator 
sequence.\cite{Sandvik_Book_2010} The system size can again be used 
to effect a scaling collapse of $\rho_s$ to the form employed in 
Fig.~\ref{fig:3},\cite{Sandvik_Book_2010,Weber_PRB_1988}  
\begin{equation}
\rho_s = \frac{2t_{\mathrm{KT}}}{\pi} \left[ 1 + \frac{1}{2 \ln L + c} \right] 
F_{\rho} \big( \ln L - a /\sqrt{t - t_{\mathrm{KT}}} \big), 
\end{equation}
where $a$ and $c$ are fitting parameters and $F_{\rho}(t,L)$ is a single 
scaling function. This collapse allows an accurate determination of the 
BKT transition temperature, $t_{\mathrm{KT}}$. 

\bibliographystyle{naturemag}

\begin{thebibliography}{10}
\bibitem{footnote1}
\bibinfo{title}{{The DTAF breaks a continuous U(1) symmetry, but retains a
  double degeneracy when the orientation of the staggered order parameter is
  defined. One configuration is illustrated in Fig.~\ref{fig:1}b and the other
  can be obtained by exchanging both spins on alternating sites of the square
  lattice (i.e.~by a sublattice transformation).}}
\end{thebibliography}

\phantom{.}

\noindent
{\bf Data availability}
The data that support the findings of this study are available 
from the corresponding author upon reasonable request.

\smallskip
\noindent
{\bf Code availability}
The code that supports the findings of this study is available 
from the corresponding author upon reasonable request.

\smallskip
\noindent
{\bf Acknowledgements}

\noindent
We thank F. Mila, Y. Wan, Y. Wang, S. Wessel and W. Yu for helpful 
discussions. This work was supported in part by the National Natural 
Science Foundation of China under Grant Nos.~11974396, 12174441 and 
12188101. We acknowledge the Physical Laboratory of High-Performance 
Computing at Renmin University of China for the provision of computational 
resources and the National Supercomputer Center in Guangzhou for the use 
of the Tianhe-2 supercomputer. 

\smallskip
\noindent
{\bf Author contributions}

\noindent
The project was conceived by B.N. and R.Y. QMC calculations were performed 
by Y.F. and PESS calculations by N.X. Theoretical interpretation was provided 
by Y.F., C.L., B.N. and R.Y. The manuscript was written by Y.F., B.N. and R.Y. 
with contributions from all the authors.

\smallskip
\noindent
{\bf Competing financial interests} The authors declare no competing 
financial interests.

\smallskip
\noindent
{\bf Additional information}

\noindent
{\bf Supplementary information} The online version contains supplementary 
material available at https://doi.org/xxx.yyy.zzz.

\noindent
{\bf Correspondence and requests for materials} should be addressed to B.N. 
and R.Y.

\clearpage

\setcounter{figure}{0}
\renewcommand{\thefigure}{S\arabic{figure}}

\setcounter{section}{0}
\renewcommand{\thesection}{S\arabic{section}}

\setcounter{equation}{0}
\renewcommand{\theequation}{S\arabic{equation}}

\setcounter{table}{0}
\renewcommand{\thetable}{S\arabic{table}}

\onecolumngrid

\noindent
{\large {\bf {Supplementary information to accompany the article}}}

\vskip4mm

\noindent
{\large {\bf {``Emergent criticality in fully frustrated quantum magnets''}}}

\vskip4mm

\noindent
Yuchen Fan, Ning Xi, Changle Liu, Bruce Normand and Rong Yu

\vskip6mm

\section{Tensor-network results for the ground states of the quantum model}

\twocolumngrid

\begin{figure*}[h]
\centering
\includegraphics[width=0.7\textwidth]{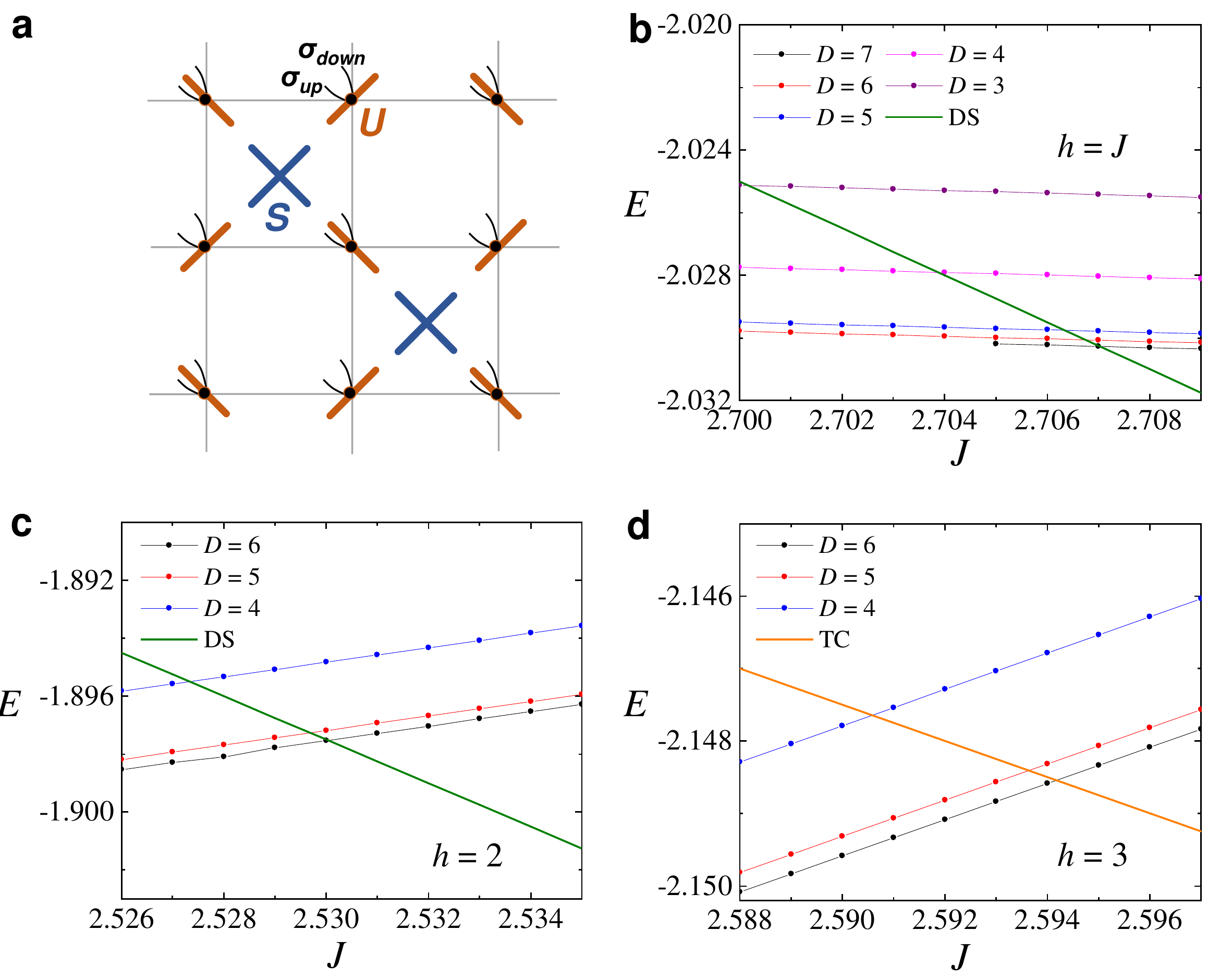}
\caption{{\bf Tensor-network calculations.} {\bf a} iPESS framework for 
bilayer computations. A simplex $S$ tensor (blue) is placed at the centres of 
alternating squares on a square lattice, and is connected to four $U$ tensors 
(red) placed on the lattice sites. Each $U$ tensor encodes the two physical 
$S = 1/2$ spins of the bilayer dimer unit and two auxiliary variables, while 
each $S$ tensor encodes four auxiliary variables. Each tensor has bond 
dimension $D$, which is the truncation parameter of the calculation. 
To draw the quantum phase diagram shown in Fig.~1c of the main text, 
iPESS is required only to compute the energy of the DTAF phase, which 
is then compared with the exactly known energies of the DS, TC and FM phases. 
{\bf b} Comparison of energies in the vicinity of first-order transitions 
between the DTAF phase (lines with symbols) and the DS phase (solid green 
line) at $h = 2$. 
{\bf c} Comparison of energies between the DTAF phase and the TC phase 
(solid orange line) at $h = 3$. 
{\bf d} Comparison of energies between the DTAF phase and the degenerate DS 
and TC phases (solid green line) at $h = J$, shown around their triple point.} 
\label{Sfig:1}
\end{figure*}

We determine the $t = 0$ phase diagram of the fully frustrated bilayer (FFB) 
model, shown in Fig.~1c of the main text, by applying the tensor-network 
method of infinite projected entangled simplex states (iPESS).
\cite{Xie_PRX_2014} In this construction, illustrated in Fig.~\ref{Sfig:1}a, 
a projection tensor, $U$, is defined on each site of a square lattice, which 
contains the two spins of the inter-layer dimer. Four $U$ tensors around a 
single square are connected by simplex tensors, $S$, which reside at the 
centres of alternating squares. The $S$ tensors do not carry any physical 
spin degrees of freedom, but are introduced to allow an optimal account of 
the spin entanglement. It has been shown that this construction delivers an 
excellent description of the ground states of a number of frustrated spin 
systems.\cite{Xie_PRX_2014,Liao_PRL_2017} iPESS calculations are performed 
for an infinitely large system by employing the translational symmetry. For 
the FFB we take a 2$\times$2 unit cell containing two independent $S$ 
tensors and four $U$ tensors. 

\onecolumngrid

\twocolumngrid

To determine the ground states, we start from a random initial state and 
perform imaginary-time evolution\cite{Jiang_PRL_2008, Xie_PRX_2014, 
Phien_PRB_2015, Wang_arXiv_2011} by the simple-update 
method.\cite{Jiang_PRL_2008, Xie_PRX_2014} After convergence of the state 
has been achieved, we use the corner-transfer-matrix renormalization-group 
(CTMRG) method\cite{Corboz_PRL_2014} to contract the infinite network and 
calculate the expectation values of physical observables. In certain 
parameter regimes, more than one converged state can be stabilized when 
starting from different initial states, and in this case the ground state 
is selected by comparing energies. From this procedure we identify four 
ground states, the dimer singlet (DS), triplon crystal (TC), dimer triplet 
antiferromagnetic (DTAF) and ferromagnetic (FM) states. The DS, TC and FM 
states are all product states, which are easily constructed in the basis 
of interlayer dimers as 
\begin{align}
 & |\mathrm{DS} \rangle = \Pi_{i=1}^{N} |S \rangle_i, \\
 & |\mathrm{TC} \rangle = \Pi_{i \in A, j \in B} |S\rangle_i |T_{+1} 
     \rangle_j, \\
 & |\mathrm{FM} \rangle = \Pi_{i=1}^{N} |T_z\rangle_i,
\end{align}
where
\begin{align}
& |S \rangle = {\textstyle \frac{1}{\sqrt{2}}} |\uparrow_1 \downarrow_2
 - \downarrow_1 \uparrow_2 \rangle \;\; {\rm and} \\
& |T_{+1} \rangle = |\uparrow_1 \uparrow_2 \rangle 
\end{align}
are dimer singlet and triplet states defined on one site of the square 
lattice. The energies of these states are readily obtained analytically as 
\begin{align}
& E_{\mathrm{DS}} = - {\textstyle \frac{3}{4}} J, \\
& E_{\mathrm{TC}} = - {\textstyle \frac{1}{4}} J - {\textstyle \frac12} h 
\qquad {\rm and} \\
& E_{\mathrm{FM}} = {\textstyle \frac{1}{4}} J - h 
\end{align}
per dimer. These product states can all be expressed as tensor-network 
states with bond dimension $D = 1$, and the energies calculated 
numerically in the iPESS representation match exactly with the 
analytical results. 

\begin{figure}[t]
\centering
\includegraphics[width=0.75\columnwidth]{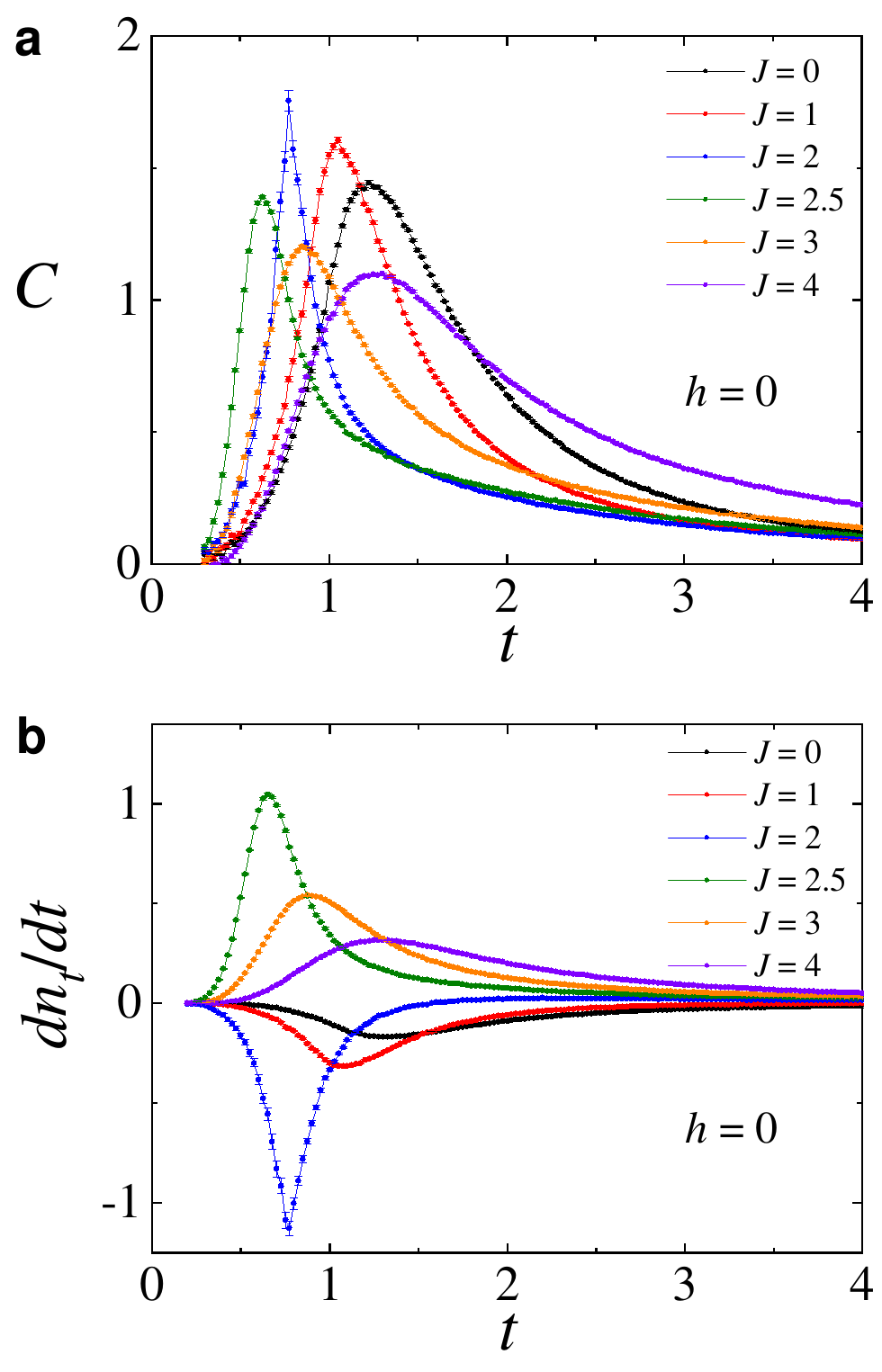}
\caption{{\bf Zero-field physics.} {\bf a} Specific heat, $C$, shown as a 
function of the reduced temperature, $t$, at $h = 0$ for selected values 
of $J$. {\bf b} Temperature derivative of the triplet density, $dn_t/dt$, 
shown as a function of $t$ for same parameters. In both cases, the peak 
position marks the characteristic temperature of the crossover to a 
correlated spin state (shown as the magenta points in Figs.~1e and 2a 
of the main text). The system size is $L = 24$.}
\label{Sfig:2}
\end{figure}

A determination of the phase boundaries therefore requires only a 
comparison of the calculated energy for the DTAF state with those of 
the other three states. Results in three representative parameter 
regimes are shown in Fig.~\ref{Sfig:1}. In each case, we find rapid 
convergence of the DTAF energy with increasing $D$, even at very moderate 
$D$ values, and thus it is not necessary to push our calculations to large 
$D$. Figures \ref{Sfig:1}b and \ref{Sfig:1}c show respectively the 
extraction of phase-boundary points for the DS-DTAF and TC-DTAF transitions. 
The DS-TC transition occurs along the line $h = J$, where the two states are 
degenerate with energy $E = - {\textstyle \frac{3}{4}} J$ per dimer, and 
the crossing point of the DTAF and DS/TC energies along this trajectory 
determines the triple point. Once again, the energy of the DTAF state 
converges rapidly with increasing $D$, and the triple point can be located 
accurately at $J = 2.707 \pm 0.001$ (using $D = 7$). Thus we find that our 
tensor-network calculations deliver precise phase boundaries for the FFB 
model, with numerical error bars arising from the finite $D$ or other 
parts of the optimization process being smaller than the symbol sizes in 
the phase diagram shown in Fig.~1c of the main text. 

\begin{figure}[t]
\includegraphics[width=0.75\columnwidth]{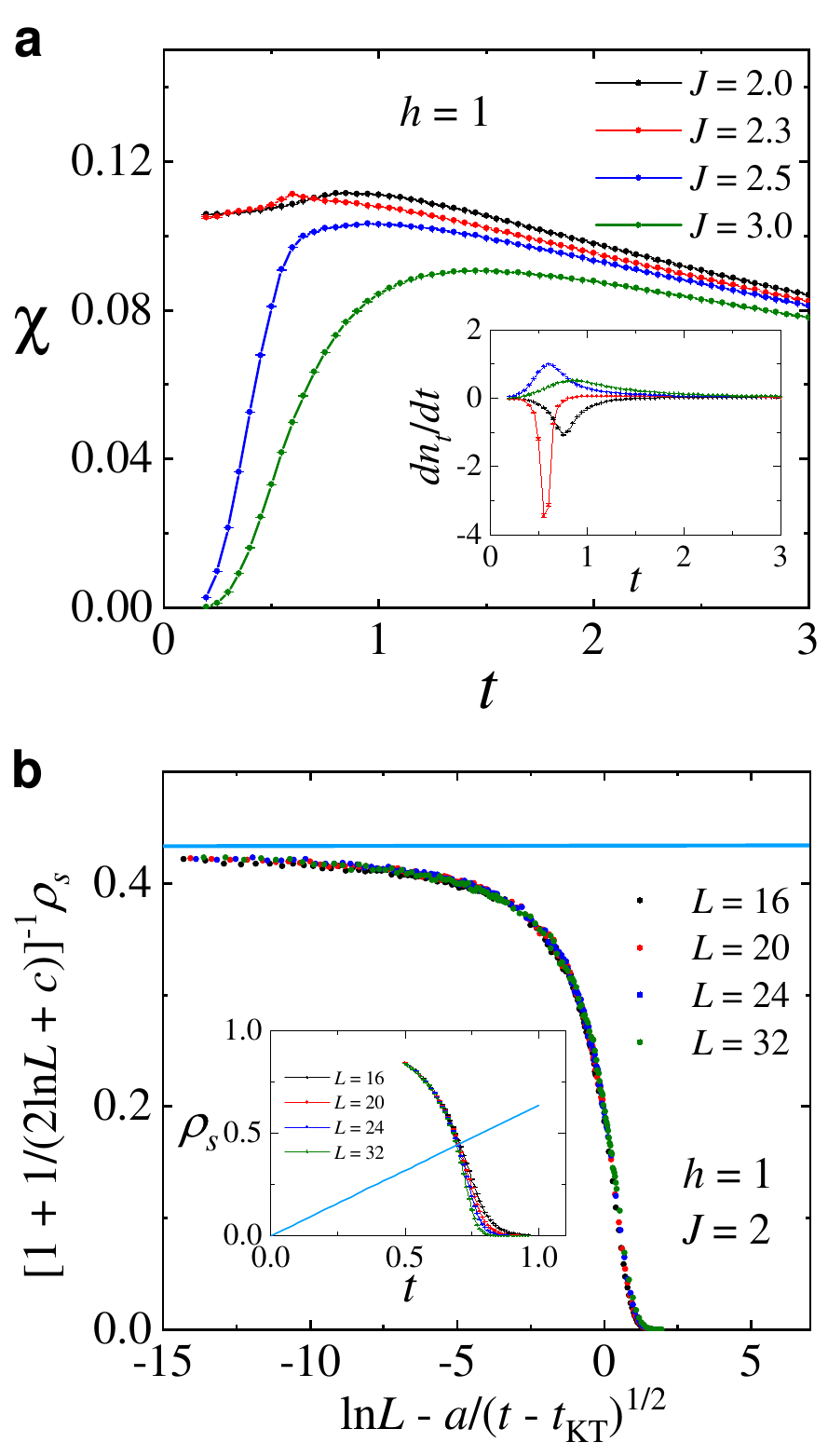}
\caption{{\bf Physics of the DTAF.} {\bf a} Uniform susceptibility, $\chi 
(t)$, computed with $L = 24$ at $h = 1$ for selected $J$ values. $\chi (t)$
converges to a finite value as $t \rightarrow 0$ in the DTAF phase. 
The inset allows a comparison of the peaks in $\chi (t)$ with those 
in $C(t)$ and $dn_t/dt$, which were used to extract the magenta points 
shown in Figs.~1e and 4b of the main text. 
{\bf b} Finite-size scaling of the spin stiffness, $\rho_s (t)$, near the 
BKT transition obtained by fitting to the form of $\rho_s(t,L)$ given in 
the Methods section of the main text, from which we determined $a = 0.8 
\pm 0.05$, $c = - 2.0 \pm 0.05$ and the BKT transition temperature 
$t_{\mathrm{KT}} = 0.676 \pm 0.002$. The inset shows $\rho_s$ calculated at 
$h = 1$ and $J = 2$ for a range of system sizes, a form in which $t_{\rm KT}$ 
is determined from the crossing point of $\rho_s (t)$ curve with the linear 
function $2t/\pi$ (cyan line) in the limit $L \rightarrow \infty$.}
\label{Sfig:3}
\end{figure}

\section{Thermal phase transitions and critical behaviour of the quantum model}

The FFB model in a field possesses a rich phase diagram at finite temperatures, 
as Fig.~1e of the main text makes clear. Although the magnetic long-range order 
of the DTAF phase cannot persist at finite temperature as a consequence of 
the Mermin-Wagner theorem, the first-order phase transitions driven by 
increasing $J$ do persist. The three ground states appearing below saturation 
have different triplet densities, with $n_t = 1$ per dimer site in the DTAF, 
$n_t = 1/2$ in the TC and $n_t = 0$ in the DS phase. At low temperatures, 
$n_t$ jumps between different parameter regimes in first-order transitions 
(shown as the walls of discontinuities in Fig.~1e), but as the temperature 
increases, the differences in triplet density reduce and eventually vanish. 
Here the walls of first-order transitions are terminated by a lines of 
critical points. Our most striking result is how the nature of this critical 
point depends on the strength of the applied magnetic field: Figs.~4a-c of 
the main text show the three different scenarios obtained when the 
three-dimensional (3D) phase diagram is analysed at different field values. 
The critical behaviour in each regime is discussed in the main text, and 
here we provide additional data that supplement all of the conclusions 
drawn there. 

\noindent
{\bf Ising critical point at $h = 0$.} The critical point terminating the 
first-order line at $h = 0$ has been studied previously.\cite{Wessel_PRL_2018} 
Here no spontaneous breaking of symmetry takes place and the first-order 
transition is of the liquid-gas type, with the critical point having Ising 
universality. The crossover temperatures shown in Fig.~4a of the main text 
were determined for both the DTAF and DS phases from the positions of the 
peaks in the specific heat $C(t)$, as shown in Fig.~\ref{Sfig:2}a. The 
crossover can also be characterized using the peak in the derivative of the 
triplet density, $dn_t/dt$, shown in Fig.~\ref{Sfig:2}b, which reflects the 
singlet-triplet nature of this transition.

\begin{figure}[t]
\centering
\includegraphics[width=0.82\columnwidth]{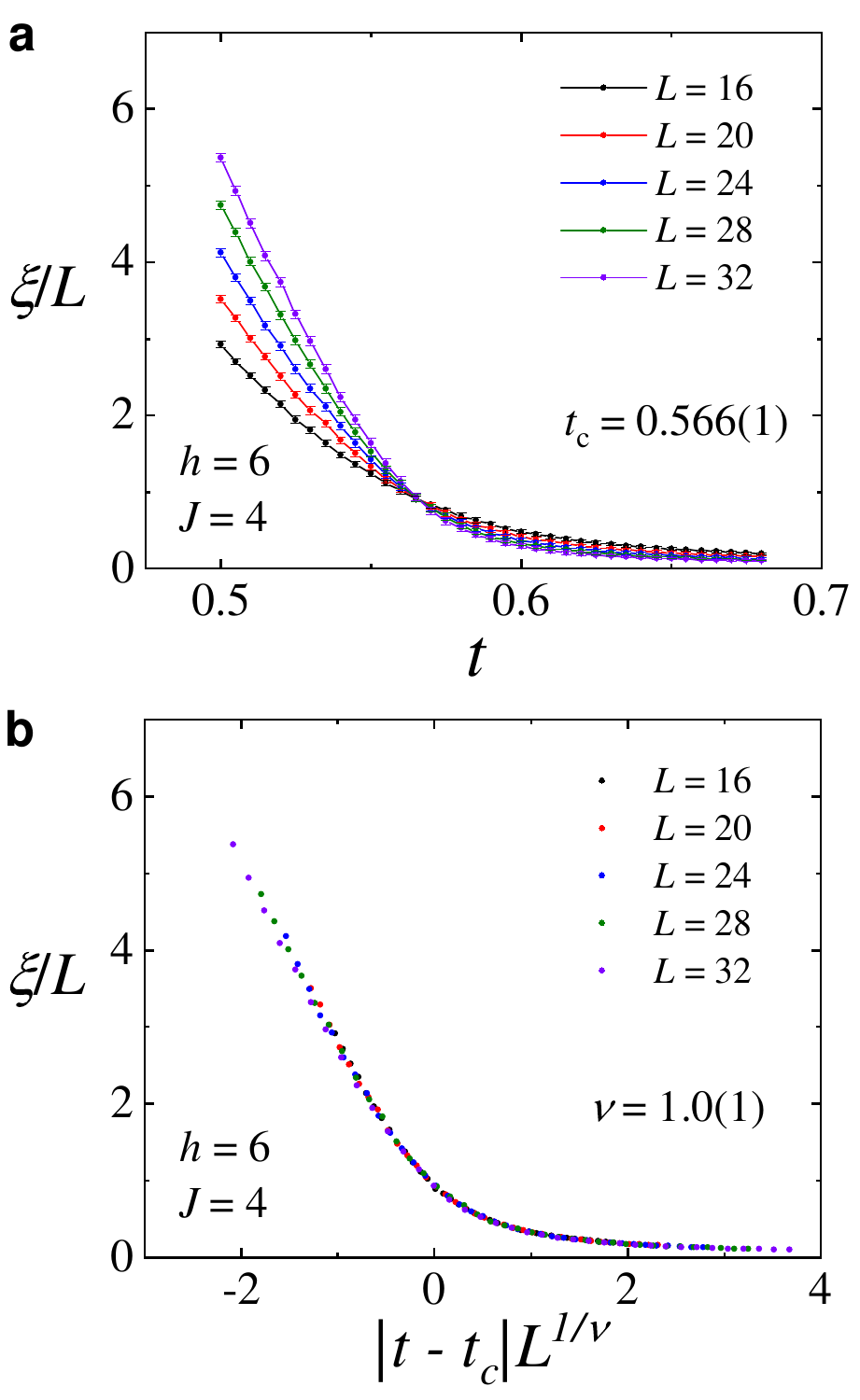}
\caption{{\bf Physics of the TC.} 
{\bf a} Correlation length, $\xi(t)$, computed for a range of system sizes 
in the TC regime ($J = 4$, $h = 6$). The transition temperature, $t_c = 0.566 
\pm 0.001$, is determined from the crossing point. {\bf b} Scaling collapse 
of $\xi (t,L)$, yielding the critical exponent $\nu = 1.0 \pm 0.1$.}
\label{Sfig:4}
\end{figure}

\noindent
{\bf Ising critical point at low fields.} At any finite field, the DTAF 
phase has only quasi-long-ranged order (qLRO) at finite temperatures, 
and one may question how this affects the nature of the critical point 
terminating the DTAF-DS transition. The Berezinskii-Kosterliz-Thouless 
(BKT) transition into the qLRO phase does not result in the spontaneous 
breaking of any symmetry, and the transition temperature, $t_{\rm{KT}}$, does 
not match with the crossover temperature characterizing the thermodynamic 
properties, as shown in Fig.~3a of the main text. In Fig.~\ref{Sfig:3}a we 
contrast the susceptibility, $\chi(t)$, in the gapped DS phase, where it 
vanishes as $t \rightarrow 0$, with that in DTAF phase, where it approaches 
a finite value due to the gapless ground state and incipient LRO. In 
Fig.~\ref{Sfig:3}b we examine the BKT transition in more detail for one 
parameter choice in the low-field regime, $h = 1$ and $J = 2$: as in 
Figs.~3b and 3c of the main text, which were computed at $h = 6$, we use 
the scaling form given in Eq.~(9)\cite{Sandvik_Book_2010,Weber_PRB_1988} 
to determine the transition temperature to high accuracy (here 
$t_{\mathrm{KT}} = 0.676 \pm 0.002$). 

\begin{figure}[t]
\centering
\includegraphics[width=0.82\columnwidth]{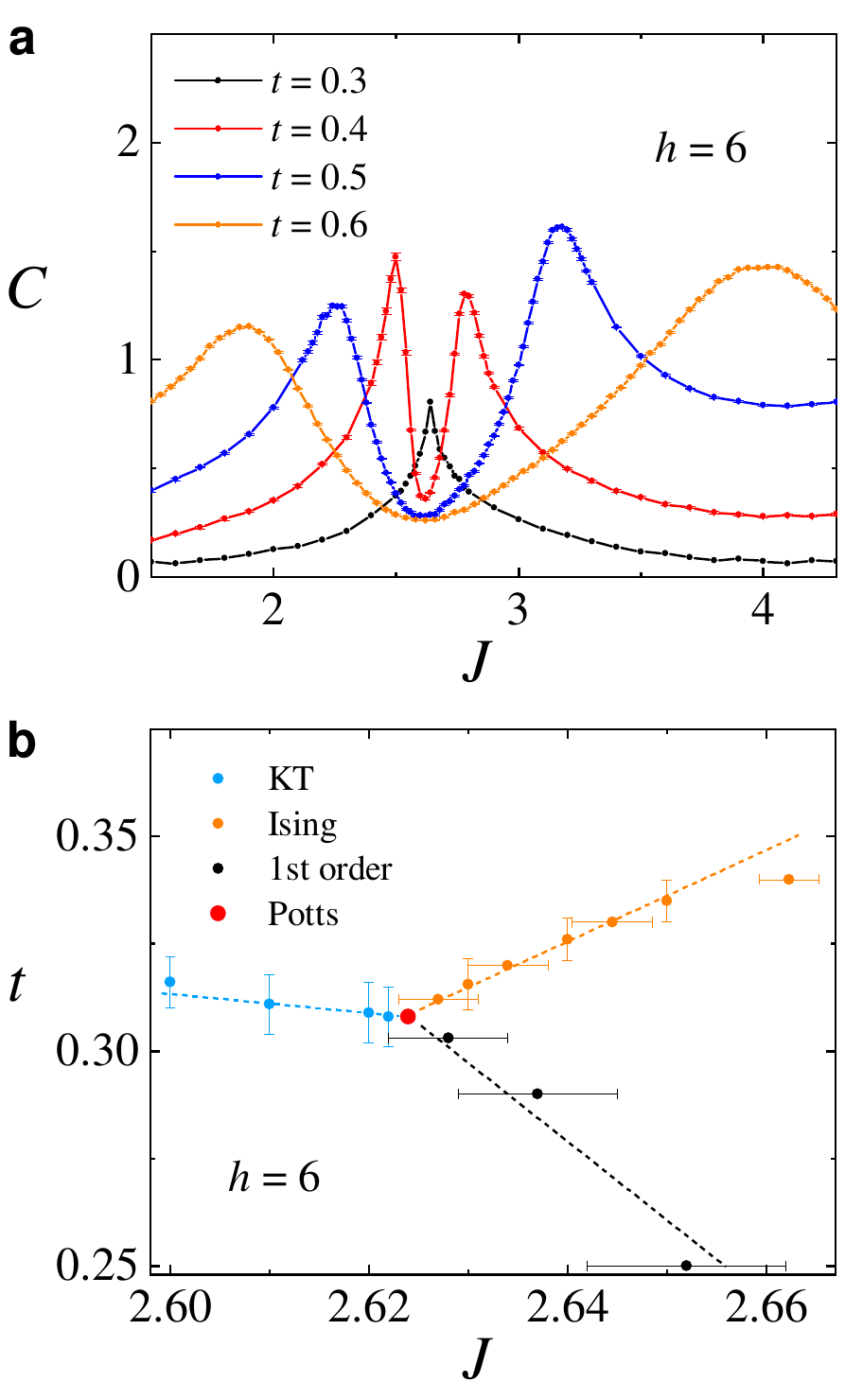}
\caption{{\bf Multicritical point.} {\bf a} Specific heat computed at $L = 24$ 
and shown as a function of $J$ at $h = 6$. The low-$J$ peaks characterize the 
crossover on the DTAF side, shown by the magenta points in Figs.~1e and 4c of 
the main text. The high-$J$ peaks characterize the TC phase. The single peak 
for $t < t_c \simeq 0.31$ indicates the DTAF-TC transition. 
{\bf b} Convergence of first-order and continuous transition lines to a single 
multicritical point, illustrated for $h = 6$. Our calculations indicate 
convergence to the single point, $(J_c, t_c) = 2.624(4), 0.308(5))$, with 
the accuracy indicated; the maximum system size employed was $L = 32$.} 
\label{Sfig:4b}
\end{figure}

At $h = 1$ we identified the critical point as $J_c = 2.36 \pm 0.01$ and 
$t_c = 0.47 \pm 0.01$. Here the two crossover lines, the BKT transition line 
and the first-order singlet-triplet transition line all meet, as shown in 
Fig.~4b of the main text. As noted there, we have demonstrated this meeting 
at the 0.01 accuracy level in our simulations; in fact the exact behaviour 
of the crossover lines\cite{Weber_SP_2022} is immaterial for the discussion 
of criticality, and we found no evidence that the BKT line might not meet 
the critical point at any $h$ value. To deduce the universality of this 
critical line, in Fig.~4d of the main text we showed that the finite-size 
scaling of the peak value of the specific heat, $C_{\mathrm{max}} \propto \ln 
L$, confirms that the critical point remains of Ising type, meaning that the 
presence of the qLRO BKT phase does not change the universality from the 
$h = 0$ case. In Sec.~S3E below we comment in more detail on this result. 

\begin{figure*}[t]
\includegraphics[width=0.92\textwidth]{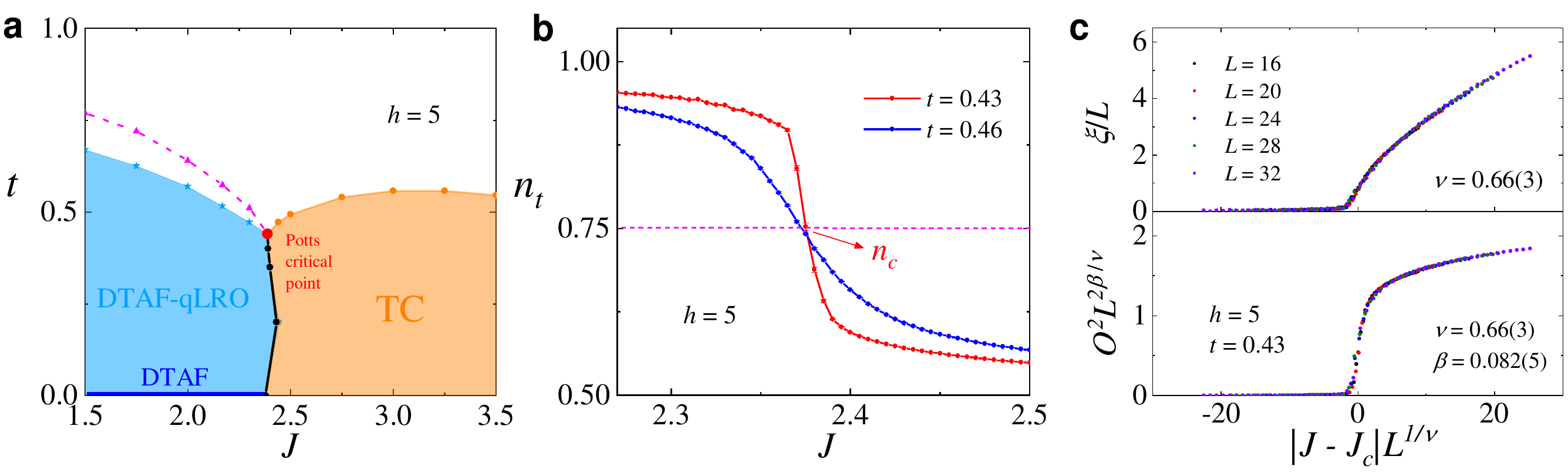}
\caption{{\bf Emergent criticality at the DTAF-TC transition.} {\bf a} Phase 
diagram of the FFB model in the $J$-$t$ plane at $h = 5$. The phase diagram 
has the same form as at $h = 6$ (Fig.~4c of the main text), with the emergent 
critical point located at $(J_c,t_c) = (2.378 \pm 0.002, 0.43 \pm 0.01)$. 
{\bf b} Triplet density, $n_t$, calculated with $L = 24$ and shown as a 
function of $J$ near the emergent critical point, where $n_t = 0.75$, 
indicating a triplet-singlet particle-hole symmetry. 
{\bf c} Scaling collapse of the correlation length, $\xi$, and the order 
parameter, $O$, of the TC phase across the emergent critical point. The 
critical exponents estimated from these two types of data collapse are 
$\nu = 0.66 \pm 0.03$ and $\beta = 0.082 \pm 0.005$, fully consistent 
with 4-state Potts universality.}
\label{Sfig:5}
\end{figure*}

\begin{table*}
\centering
\begin{tabular}{|c|c|c|c|c|c|c|}
\hline
$\; \vec{T} \;$ & $|S \rangle$ & $|T_{-1} \rangle$ & $|T_{0} \rangle$ & 
$|T_{+1} \rangle$ & $\;$unphysical$\;$ & $\;$unphysical$\;$ \\ \hline
$n$ & $0$ & $1$ & $1$ & $1$ & $0$ & $0$ \\ \hline
$\mathbf{S}$ & $\; |S^z = 0 \rangle \;$ & $\; |S^z = - 1 \rangle \;$ & 
$\; |S^z = 0 \rangle \;$ & $\; |S^z = 1 \rangle \;$ & $|S^z = - 1 
\rangle$ & $|S^z = 1 \rangle$ \\ \hline
\end{tabular}
\caption{Mapping between the dimer basis (top row) and the slave-spin 
representation (centre and bottom rows). The unphysical states are 
eliminated by enforcing a constraint in the slave-spin representation.}
\end{table*}

\noindent
{\bf Emergent critical point at intermediate fields.} When $h > 
h_{\mathrm{tr}}$, the triple point where the DS, TC and DTAF phases meet, 
the wall of discontinuities separates the DTAF ($n_t = 1$) from the TC 
($n_t = 0.5$), where the alternation of dimer singlets and triplets breaks 
the discrete lattice (translational) symmetry. Thus, as shown in the main 
text, TC order persists at finite temperatures, exhibiting a continuous 
thermal transition to the disordered (paramagnetic) phase. In 
Fig.~\ref{Sfig:4}a we show the correlation length, $\xi(t)$, calculated 
for different system sizes at one point in the TC phase ($h = 6$ and $J
 = 4$). From the crossing point of the $\xi(t,L)$ lines we determine the 
transition temperatures to high accuracy (here $t_c = 0.566 \pm 0.001$). 
In Fig.~\ref{Sfig:4}b we implement the scaling collapse of our finite-size 
data for $\xi$ that is analogous to the collapse of the order parameter, 
$O$, of the TC phase (the staggered longitudinal magnetization) shown in 
Fig.~2c of the main text. Both $\xi$ and $O$ confirm that the thermal 
transition has Ising universality, with critical exponents $\nu = 1$ and 
$\beta = 1/8$.

In Fig.~\ref{Sfig:4b}a we use the specific heat at fixed $h = 6$ to 
characterize the convergence of the DTAF crossover temperature and the 
TC transition temperature. Where the two meet as the temperature is reduced 
marks the critical point, and below this temperature the transition becomes 
first-order. At the DTAF-TC transition, the issue of the accuracy with which 
we can demonstrate convergence to one multicritical point becomes important, 
because even without the crossover lines this point should mark the meeting 
of one first-order transition line with two well defined lines of continuous 
transitions (one of which is the KT transition to a qLRO phase). In 
Fig.~\ref{Sfig:4b}b we quantify the statement that such a meeting is 
demonstated directly at the 0.01 accuracy level. As noted in the main text, 
our simulations have a finite accuracy and thus cannot demonstrate exact 
convergence; instead we base our analysis of the multicritical point on 
(i) the application of symmetry arguments to the equivalent classical model, 
(ii) its immediate connection to the well accepted bicritical-point scenario 
if the DTAF phase is given any finite spin anisotropy and (iii) the high 
precision to which we obtain the scaling exponenents of a qualitatively 
different universality class at the multicritical point estimated to the 
available precision. 

To reiterate the last point, our most remarkable result is the emergent 
symmetry of the multicritical point (i.e.~the point terminating the first-order 
DTAF-TC transition). As we showed in Fig.~4f of the main text, the scaling of 
the TC order parameter and the corresponding correlation length through this 
point give to high accuracy the critical exponents $\nu = 2/3$ and $\beta = 
1/12$ of 4-state Potts universality.\cite{Potts_RMP_1982} To demonstrate that 
this universality class is a global property of the full line of emergent 
critical points, i.e.~over the field regime $h_{\rm{tr}} < h < h_{\rm{QCEP}}$, 
we have calculated the scaling properties at additional field values and 
in Fig.~\ref{Sfig:5} we show our results for $h = 5$. 

Figure \ref{Sfig:5}a shows that the fixed-field phase diagram is very 
similar to the $h = 6$ case shown in Fig.~4c of the main text. Figure 
\ref{Sfig:5}b shows the evolution of $n_t$ as $J$ is varied at temperatures 
at and above $t_c$. At the critical point, $n_t = 3/4$, as also shown in 
Fig.~4e of the main text. The DTAF and the TC have two different types of 
order that both break the Z$_2$ sublattice symmetry, giving four ground-state 
configurations that are degenerate at the $t = 0$ transition, which has an 
exact triplet-singlet particle-hole symmetry about $n_t = 3/4$. Our results 
imply that this particle-hole symmetry persists at finite temperature, and 
indeed up to the emergent critical point. The importance of this symmetry in 
understanding the nature of the emergent Potts universality is detailed in 
Sec.~S3.
 
Figure \ref{Sfig:5}c shows the scaling collapse of the correlation length 
and order parameter, from which we again obtain the critical exponents of 
the 4-state Potts model to the same precision. We comment that the accuracy 
of our calculations depends on having a robust $t_c$, 
and thus our error bars become larger as the field is increased well beyond 
$h = 6$, where $t_c$ decreases monotonically (Fig.~1e of the main text). 
Again one may ask how this emergent Potts universality arises from the 
symmetries of the DTAF and TC phases, and whether it is influenced by the 
BKT physics of the former. Next we study the classical, spin-anisotropic 
and Ashkin-Teller models offering insight into these questions.

\begin{figure*}[t]
\includegraphics[width=0.96\textwidth]{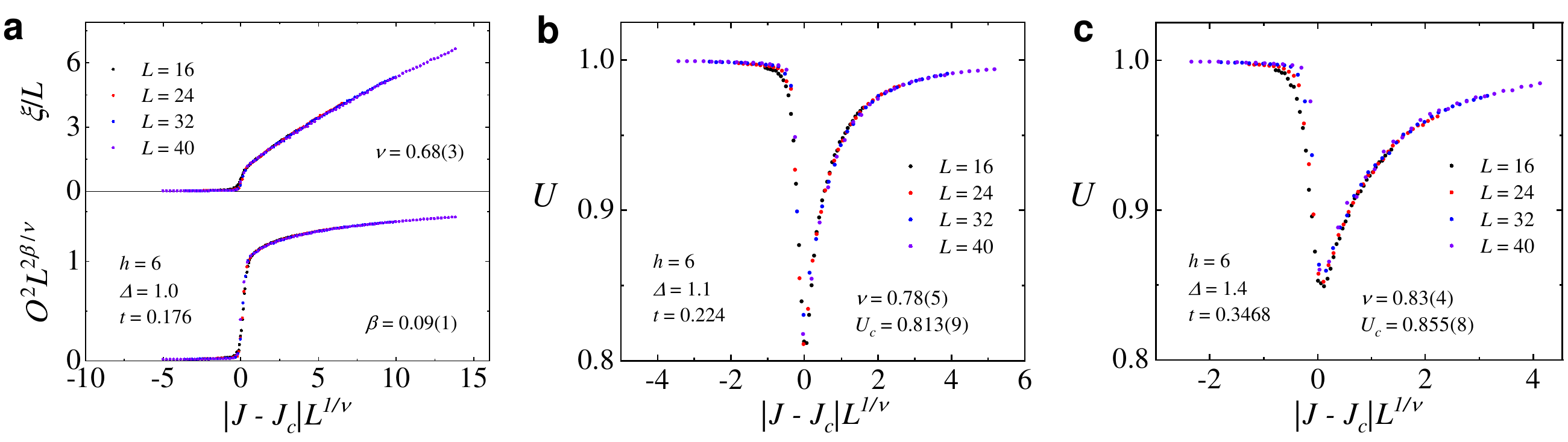}
\caption{{\bf Emergent criticality in the classical and spin-anisotropic 
models.} {\bf a} Scaling collapse of the correlation length and order 
parameter of the TC phase of the classical model at $h = 6$, shown as a 
function of $|J - J_c|$ at fixed $t_c$. The emergent critical point is 
located at $(J_c, t_c) = (1.917 \pm 0.004, 0.176 \pm 0.002)$ and the 
critical exponents estimated from these two panels are $\nu = 0.68 \pm 
0.03$ and $\beta = 0.09 \pm 0.01$.
{\bf b} Collapse of Binder-cumulant data shown for the spin-anistropic 
model at $\Delta = 1.1$. {\bf c} Collapse of Binder-cumulant data shown 
for the spin-anistropic model at $\Delta = 1.4$.} 
\label{Sfig:6}
\end{figure*}

\section{Emergent criticality in a classical model}

\subsection{Mapping to a classical model}

To better understand the nature of the emergent critical point, and to 
allow more efficient simulations, we map the quantum FFB model in the 
dimer basis [Eq.~(2) of the main text] to a classical model defined on 
the square lattice. We begin by writing
\begin{equation}
\vec{T}_i = n_i \mathbf{S}_i 
\end{equation}
at each lattice site, where $n_i = 0$ or $1$ is an Ising variable that 
encodes the triplet density and $\mathbf{S}_i$ is a slave spin-1 operator. 
In Table~S1 we present the mapping between the local states in the dimer 
basis and the slave-spin representation, which has an enlarged Hilbert 
space. The quantum Hamiltonian is then mapped, up to a constant, to 
\begin{eqnarray}
\label{Eq:HamCl}
H_{\rm{cl}} & = & \sum_{\langle i,j \rangle} n_i n_j \mathbf{S}_i \cdot 
\mathbf{S}_j + J\sum_i n_i^2 \nonumber \\ & & \quad - h\sum_i n_i S^z_i
 + D \sum_i (1 - n_i) \left( S^z_i \right)^2, 
\end{eqnarray}
where the last term is the constraint. The unphysical states are projected 
out completely in the limit $D \rightarrow + \infty$, but for calculational 
purposes we employ a soft constraint with $D \gg J$ and $h$.

To study the phase transitions at finite temperatures near the emergent 
critical point, we make a further approximation by treating $\mathbf{S}_i$ 
as a classical O(3) rotor with unit magnitude. This turns Eq.~\eqref{Eq:HamCl} 
into a classical model of Ising variables coupled to O(3) rotors on the square 
lattice, with the Zeeman coupling between the magnetic field and the rotor 
reducing the spin rotational symmetry to an in-plane O(2).

\subsection{Phase diagram and emergent criticality}

This classical model has a global phase diagram very similar to that of 
the quantum model, as we showed in Fig.~5a of the main text. Ordering of 
either the Ising or the rotor variables returns all four phases of the 
quantum model: the magnetic phases, DTAF and FM, correspond to ordering 
of the O(3) rotors, while the DS state has the Ising variable $n_i
 = 0$ (singlets) at every site and the TC state is expressed by an 
alternating order of the Ising variables ($n_i = 0$ and $n_i = 1$) on 
the two sublattices. We note that an advantage of this classical model 
is that it allows the transition to be studied by explicit control of 
the density ($n$). 

Going beyond the phase diagrams (Figs.~4c and 5a of the main text), in 
Fig.~\ref{Sfig:6}a we perform the same scaling-collapse analysis for the 
correlation length and order parameter of the TC phase that we performed in 
Fig.~4f of the main text, albeit with larger system sizes. Somewhat remarkably, 
the critical exponents once again point to emergent 4-state Potts universality 
in the classical approximation, being entirely consistent with the values 
$\nu = 2/3$ and $\beta = 1/12$. This property of the classical model allows us 
to reveal the nature of the critical point by two further pieces of analysis.

\subsection{Classical model with spin anisotropy}

Next we introduce a spin anisotropy in the classical model, so that 
Eq.~\eqref{Eq:HamCl} becomes
\begin{eqnarray}
H_{\rm{cl}} & = & \sum_{\langle i,j \rangle} n_i n_j \left(\Delta S^x_i 
S^x_j + S^y_i S^y_j + S^z_i S^z_j \right) + J \sum_i n_i^2 \nonumber \\
& & \quad - h\sum_i n_i S^z_i + D\sum_i (1 - n_i) \left( S^z_i \right)^2. 
\label{Eq:HamIsing}
\end{eqnarray}
For $\Delta > 1$, the spin anisotropy is of Ising type and the BKT 
transition is replaced by an Ising transition into a phase with long-ranged 
AF order. The anisotropy has no effect on the TC phase, or on the first-order 
transition into it, as we showed in the phase diagram of Fig.~5b of the main 
text. In this model [Eq.~\eqref{Eq:HamIsing}], the critical point where the 
continuous x-Ising AFM and TC transition lines meet is a conventional 
bicritical point, below which the direct AFM-TC transition is first-order.  

\begin{table*}[t]
\centering
\begin{tabular}{|c|cccc|}
\hline
$|i \rangle$ & $|1 \rangle$ & $|2 \rangle$ & $|3 \rangle$ & $|4 \rangle$ 
\\ \hline $\; |\sigma, \eta, \tau \rangle \;$ & $\; |- 1, + 1, - 1 \rangle$ 
& $|+ 1, + 1, + 1 \rangle$ & $|- 1, - 1, + 1 \rangle$ & $|+ 1, - 1, - 1 
\rangle \;$ \\ \hline
\end{tabular}
\caption{Sets of Ising variables labelling the degenerate ground-state 
configurations of the effective classical model.}
\end{table*}

Studying the critical properties of this model is complicated by the 
fact that both phases have long-range order and both break the same 
sublattice symmetry. For a reliable analysis we turn to the Binder 
cumulant,\cite{Goldenfeld_Book_1992,Dohm_PRE_2004,Selke_JPA_2005,
Selke_EPJB_2006,Songbo_PRL_2012,Songbo_PRB_2013} and here we provide full 
details of the approach summarized in the main text. We define a composite 
order parameter, $O_b^2 = O_{\rm x}^2 + O_{\rm TC}^2$, that contains the order 
of both 
the AFM and TC phases. 
\begin{equation}
O_{\rm x} = |m_A - m_B|, \; {\rm with} \;\; m_{A/B} = \frac{1}{N} \!
\sum_{i \in A/B} \! S^x_i/|S^x_i|, \nonumber
\end{equation}
is the normalized order parameter of the x-Ising AFM phase and
\begin{equation}
O_{\rm TC} = |n_A - n_B|, \; {\rm with} \;\; n_{A/B} = \frac{2}{N} \!
\sum_{i \in A/B} \! n_i, \nonumber
\end{equation}
is the order parameter of the TC phase. We then define the Binder 
cumulant associated with $O_b$ as 
\begin{equation}
U = 2 \left( 1 - \frac{\langle O_b^4 \rangle}{2\langle O_b^2 \rangle^2} 
\right) \!. 
\end{equation} 
In Fig.~5d of the main text we showed how analysing $U(J)$ for a range 
of system sizes allows an accurate determination of the location of the 
bicritical point for a fixed value of $\Delta$, as well as the value $U_c$ 
there. In Fig.~5e we used the scaling collapse of $U$ as another means of 
extracting $\nu$, and thereby demonstrating that the critical exponents 
take non-universal values. In Fig.~\ref{Sfig:6}b we show another example 
of the scaling collapse of $U$, prepared for a weak spin anisotropy of 
$\Delta = 1.1$, and in Fig.~\ref{Sfig:6}c we show the stronger spin 
anisotropy $\Delta = 1.4$. Taken together with Figs.~5d and 5e of the main 
text, these data form the basis of Fig.~5f, where we used $U_c$ to show that 
the non-universal behaviour has a direct (linear) dependence on $\Delta$, 
reinforcing the expectation that the criticality of the model changes 
continuously between Ising and 4-state Potts universality. For an analytical 
understanding of this result we appealed to the fact that the Ashkin-Teller 
model (ATM) displays the same type of physics. 

\subsection{Ashkin-Teller model}
\label{Sec:ATM}

Interpreting the non-universal behaviour of the bicritical point yields
considerable insight into the emergent critical point. The ATM, defined 
in Eq.~(3) of the main text, has non-universal critical exponents along 
a line of critical points as the model parameter ratio $K/J$ is 
varied.\cite{Wiseman_PRE_1993} To examine the connection to the ATM, we 
map the anisotropic classical model of Eq.~\eqref{Eq:HamIsing} to an 
effective model valid around the transition where the x-Ising AFM and 
TC ground states are degenerate. Within each two-site unit cell one has 
two degenerate AFM ground states, $|1 \rangle = |\cos \theta \, T^z - \sin 
\theta \, T^x \rangle_{A} |\cos \theta \, T^z + \sin \theta \, T^x 
\rangle_{B}$ and $|2 \rangle = |\cos \theta \, T^z + \sin \theta \, T^x 
\rangle_{A} |\cos \theta \, T^z - \sin \theta \, T^x \rangle_{B}$, and two 
degenerate TC ground states, $|3 \rangle = |s \rangle_{A} |T^z \rangle_{B}$ 
and $|4 \rangle = |T^z \rangle_{A} |s \rangle_{B}$, where $|s \rangle$ 
denotes a singlet state and $T^x$ and $T^z$ are components of the triplet 
operator. 

These four ground-state configurations can be labelled by a set of three 
Ising variables, $\sigma$, $\eta$ and $\tau$, each of which takes the 
values $\pm 1$. The variable $\sigma$ is associated with the sublattice 
symmetry $A \leftrightarrow B$, so that $\sigma \propto (T^x_A - T^x_B)$ 
in the AF state and $\sigma \propto (T^z_A - T^z_B)$ in the TC state. The 
variable $\eta \propto {\vec{T}}^2_A + {\vec{T}}^2_B - 3$ is associated with 
the particle-hole symmetry about the triplet density $n_t = 3/4$ per site of 
the square lattice, which arises when the relevant low-energy excitations 
are those associated with the AFM and TC phases. The third variable is 
$\tau \equiv \sigma \eta$, meaning that only any two out of the set 
$\{ \sigma, \eta, \tau \}$ are independent. The labelling of the states 
$|i \rangle$ with the variables $|\sigma, \eta, \tau \rangle$ is shown 
in Table~S2. 

These definitions allow us to construct the effective model. The 
sublattice symmetry requires this model to be invariant under $\sigma 
\rightarrow - \sigma$ and the particle-hole symmetry requires it to be 
invariant under $\eta \rightarrow - \eta$. A minimal model with these 
symmetries is
\begin{equation}
\label{Eq:Hameff0}
H_{\rm{eff}} = - J_{\sigma} \sum_{i,j} \sigma_i \sigma_j - J_{\eta} 
\sum_{i,j} \eta_i \eta_j - J_{\tau} \sum_{i,j} \tau_i \tau_j.
\end{equation}
However, the original model is also invariant under the spin inversion 
$T^x \rightarrow - T^x$, which exchanges $|1 \rangle$ and $|2 \rangle$ 
while leaving $|3 \rangle$ and $|4 \rangle$ unchanged. In terms of 
$\sigma$, $\eta$ and $\tau$, this corresponds to $\sigma \rightarrow
 - \tau$ and $\tau \rightarrow - \sigma$, and enforcing this symmetry 
in Eq.~\eqref{Eq:Hameff0} leads to $J_{\sigma} = J_{\tau} = J$. One 
may also make use of the identity $\eta = \sigma (\sigma \eta) = \sigma 
\tau$, so that Eq.~\eqref{Eq:Hameff0} can be rewritten with $K = J_{\eta}$ as  
\begin{equation} 
\label{Eq:HamATM}
H_{\rm{eff}} = - J \sum_{i,j} (\sigma_i \sigma_j + \tau_i \tau_j) - K 
\sum_{i,j} (\sigma_i \tau_i \sigma_j \tau_j),
\end{equation}
which is Eq.~(3) of the main text.

Equation \eqref{Eq:HamATM} is precisely the ATM.\cite{Wiseman_PRE_1993} 
At $K = 0$, the ATM corresponds to two independent Ising models, and 
for $0 < K \leq J$ it has a single finite-temperature transition, where 
in the low-temperature ordered phase $\langle \sigma \rangle$, $\langle 
\tau \rangle$ and $\langle \sigma \tau \rangle$ are all non-zero. Most 
important from the perspective of the spin-anisotropic model is that 
when $0 < K/J < 1$ this transition has non-universal properties, 
exhibiting critical exponents that vary continuously between Ising 
and 4-state Potts universality, the latter reached at $K = J$. 

Returning now to our calculations, in Fig.~5f of the main text we compared 
the critical value, $U_c$, of the Binder cumulant at the bicritical points 
obtained for several different values of $\Delta$ in the spin-anisotropic 
model [Eq.~\eqref{Eq:HamIsing}] with the $U_c$ values at the critical points 
of ATMs with different $K/J$. When $\Delta$ and $K/J$ are scaled appropriately 
(the lower and upper axes of Fig.~5f), the $U_c$ values of the two models 
coincide. Thus we confirm our conclusion that the non-universal behaviour of 
the bicritical point in the spin-anistropic classical model can be described 
by the physics of the ATM. Finally, when the model is tuned to the isotropic 
limit ($\Delta \rightarrow 1$), the extrapolated $U_c$ we compute approaches 
the value $U_c \simeq 0.792$ of the ATM at its isotropic point ($K = 
J$),\cite{Songbo_PRL_2012,Songbo_PRB_2013} where the universality class is 
4-state Potts. Thus the ATM also describes the emergent criticality we 
observe in the classical models, and by extension in the quantum FFB model. 

\subsection{Conformal Field Theory}

This extraction of the equivalent classical and effective critical models 
also sheds light on our result that the BKT transition of the DTAF phase has 
no effect on the universality classes of the critical lines of the FFB in a 
field. Critical theories in two-dimensional models are described by CFTs, 
from which it is found\cite{Friedan_PRL_1984} that continuously varying 
critical exponents are possible when the conformal (central) charge of the 
CFT is $c \ge 1$, whereas transitions described by CFTs with $c < 1$ are 
restricted to discrete critical exponents. In a conventional scenario where 
two symmetries, described separately by CFTs with central charges $c_1$ and 
$c_2$, are combined at a critical point, one might expect the relevant CFT 
to have $c = c_1 + c_2$, as in the well known FFXY [U(1)$\otimes$Z$_2$] 
model.\cite{Cristofano_JSMTE_2006} However, the FFB model in a field turns 
out to be unconventional. 

The Ising critical point of the spin-isotropic model at $h = 0$ is described 
by a CFT with central charge $c = 1/2$. The BKT transition has XY universality, 
and hence is described by a $c = 1$ CFT (free bosons). To understand our result 
in the low-field regime ($h < h_{\mathrm{tr}}$), it is instructive to consider the 
model with a finite spin anisotropy, such that the BKT transition is replaced 
by an Ising transition to true LRO: in this model, both at $h = 0$ and at small 
finite $h$, the universality at the critical point becomes tricritical Ising, 
with $c = 7/10$. In the triplon basis, the Ising transition is driven by the 
triplet density, $T^2$, while in the presence of a spin anisotropy $T_x$ is 
also a relevant operator, coupling to $T^2$ by a term of the form $- T_x^2 T^2$. 
This coupling drives the tricritical Ising behaviour by changing the sign of 
the quartic term ($T_x^4$), and one may view the transition as being driven by 
the triplon amplitude mode ($|T_x|$). By contrast, the BKT transition is driven 
by the phase mode, which couples only marginally to $T^2$, this being invariant 
under a U(1) phase change. Thus one observes in the classical model how a BKT 
transition cannot have the same effect on the universality as a spin 
anisotropy, and instead leaves the critical properties in the same discrete 
(Ising) universality class as at $h = 0$. 

At fields high enough that the phase transition is DTAF-TC, the sublattice 
symmetry becomes an additional relevant global symmetry. As shown above, this 
is represented by a second independent Ising variable, whence the appropriate 
effective theory is the ATM. The ATM with $K \le J$ is described by a $c = 1$ 
CFT, meaning that continuously variable critical exponents are allowed, as we 
found in Fig.~5f of the main text both for the ATM and for the classical model 
with spin anisotropy. However, the fact that the BKT transition has $c = 1$ 
makes it only a marginally relevant perturbation of a $c = 1$ theory, and thus 
its presence should have no effect on the approach of these exponents to their 
4-state Potts values as $K \rightarrow J$ in the ATM and as $\Delta \rightarrow 
1$ in the classical model. We conclude that the effect of the qLRO BKT phase on 
the critical properties of the FFB in an applied field is at most marginal, 
explaining why it has no effect on the universality classes we find.

\subsection{Summary}

To summarize this analysis, we have employed the spin inversion, sublattice 
(or translational) and particle-hole (or singlet-triplet) symmetries to 
construct an effective ATM that describes the classical analogue of the FFB 
in the presence of an Ising spin anisotropy. For generic couplings, the ATM 
of Eq.~\eqref{Eq:HamATM} does indeed have Z$_2$$\times$Z$_2$$\times$Z$_2$ 
symmetry. However, when $K/J = 1$ a higher S$_4$ symmetry emerges, which allows 
a full permutation among the four degenerate ground-state configurations. It is 
easy to see that the spin-anisotropic model of Eq.~\eqref{Eq:HamIsing} has only 
the sublattice and spin-inversion symmetries (Z$_2$$\times$Z$_2$) of the 
generic ATM. In the isotropic limit, the spin-inversion symmetry is enhanced 
to a continuous O(2) (or U(1)) spin-rotational symmetry, but we stress that 
an S$_4$ symmetry does not exist either in this classical Hamiltonian 
[Eq.~\eqref{Eq:HamCl}] or in the original FFB model [Eq.~(1) of the main text]. 
It is an emergent symmetry appearing only at the critical point of the DTAF-TC 
transition, and this is why we adopt the terminology of an emergent critical 
point.

\end{document}